\documentclass[longauth]{aa} 

\usepackage[switch]{lineno} 

\usepackage{pdflscape} 

\usepackage{float}

\usepackage{bm}

\usepackage{comment}

\usepackage{multirow}

\usepackage{amsmath}

\usepackage{adjustbox}
\usepackage{array}
\usepackage{graphicx}
\usepackage{enumitem}
\usepackage{siunitx}
\usepackage{footmisc}
\usepackage{booktabs}
\usepackage[version=4]{mhchem}
\usepackage{svg}
\usepackage{soul,xcolor}
\setstcolor{red}
\usepackage{natbib}
\bibpunct{(}{)}{;}{a}{}{,} 
\usepackage{txfonts}
\usepackage[colorlinks=true,
            linkcolor=blue,
            urlcolor=blue,
            citecolor=blue]{hyperref}

\def\fsed{f_{\textrm{sed}}}

\makeatletter
\renewcommand*\aa@pageof{, page \thepage{} of \pageref*{LastPage}}
\makeatother

\begin{document}

\title{Orbit and atmosphere of HIP~99770~b through the eyes of VLTI/GRAVITY}

\titlerunning{New characterisation of HIP~99770~b}
\authorrunning{T.~O.~Winterhalder et al.}

\author{T.~O.~Winterhalder\inst{\ref{esog}}
 \and J.~Kammerer\inst{\ref{esog}}
 \and S.~Lacour\inst{\ref{lesia},\ref{esog}}
 \and A.~M\'erand\inst{\ref{esog}}
 \and M.~Nowak\inst{\ref{cam}}
 \and T.~Stolker\inst{\ref{leiden}}
 \and W.~O.~Balmer\inst{\ref{jhupa},\ref{stsci}}
 \and G.-D.~Marleau\inst{\ref{bernWP}, \ref{mpia}, \ref{duisburg}}
 \and R.~Abuter\inst{\ref{esog}}
 \and A.~Amorim\inst{\ref{lisboa},\ref{centra}}
 \and R.~Asensio-Torres\inst{\ref{mpia}}
 \and J.-P.~Berger\inst{\ref{ipag}}
 \and H.~Beust\inst{\ref{ipag}}
 \and S.~Blunt\inst{\ref{northwestern}}
 \and M.~Bonnefoy\inst{\ref{ipag}}
 \and H.~Bonnet\inst{\ref{esog}}
 \and M.~S.~Bordoni\inst{\ref{mpe}}
 \and G.~Bourdarot\inst{\ref{mpe}}
 \and W.~Brandner\inst{\ref{mpia}}
 \and F.~Cantalloube\inst{\ref{lam}}
 \and P.~Caselli \inst{\ref{mpe}}
 \and B.~Charnay\inst{\ref{lesia}}
 \and G.~Chauvin\inst{\ref{cotedazur}}
 \and A.~Chavez\inst{\ref{northwestern}}
 \and E.~Choquet\inst{\ref{lam}}
 \and V.~Christiaens\inst{\ref{liege}}
 \and Y.~Cl\'enet\inst{\ref{lesia}}
 \and V.~Coud\'e~du~Foresto\inst{\ref{lesia}}
 \and A.~Cridland\inst{\ref{leiden}}
 \and R.~Davies\inst{\ref{mpe}}
 \and R.~Dembet\inst{\ref{lesia}}
 \and J.~Dexter\inst{\ref{boulder}}
 \and A.~Drescher\inst{\ref{mpe}}
 \and G.~Duvert\inst{\ref{ipag}}
 \and A.~Eckart\inst{\ref{cologne},\ref{bonn}}
 \and F.~Eisenhauer\inst{\ref{mpe}}
 \and N.~M.~F\"orster Schreiber\inst{\ref{mpe}}
 \and P.~Garcia\inst{\ref{centra},\ref{porto}}
 \and R.~Garcia~Lopez\inst{\ref{dublin},\ref{mpia}}
 \and T.~Gardner\inst{\ref{exeterAstro}}
 \and E.~Gendron\inst{\ref{lesia}}
 \and R.~Genzel\inst{\ref{mpe},\ref{ucb}}
 \and S.~Gillessen\inst{\ref{mpe}}
 \and J.~H.~Girard\inst{\ref{stsci}}
 \and S.~Grant\inst{\ref{mpe}}
 \and X.~Haubois\inst{\ref{esoc}}
 \and G.~Hei\ss el\inst{\ref{actesa},\ref{lesia}}
 \and Th.~Henning\inst{\ref{mpia}}
 \and S.~Hinkley\inst{\ref{exeter}}
 \and S.~Hippler\inst{\ref{mpia}}
 \and M.~Houll\'e\inst{\ref{cotedazur}}
 \and Z.~Hubert\inst{\ref{ipag}}
 \and L.~Jocou\inst{\ref{ipag}}
 \and M.~Keppler\inst{\ref{mpia}}
 \and P.~Kervella\inst{\ref{lesia},\ref{french_chilean_lab}}
 \and L.~Kreidberg\inst{\ref{mpia}}
 \and N.~T.~Kurtovic\inst{\ref{mpe}}
 \and A.-M.~Lagrange\inst{\ref{ipag},\ref{lesia}}
 \and V.~Lapeyr\`ere\inst{\ref{lesia}}
 \and J.-B.~Le~Bouquin\inst{\ref{ipag}}
 \and D.~Lutz\inst{\ref{mpe}}
 \and A.-L.~Maire\inst{\ref{ipag}}
 \and F.~Mang\inst{\ref{mpe}}
 \and P.~Molli\`ere\inst{\ref{mpia}}
 \and C.~Mordasini\inst{\ref{bernWP}, \ref{bernCSH}}
 \and D.~Mouillet\inst{\ref{ipag}}
 \and E.~Nasedkin\inst{\ref{mpia}}
 \and T.~Ott\inst{\ref{mpe}}
 \and G.~P.~P.~L.~Otten\inst{\ref{sinica}}
 \and C.~Paladini\inst{\ref{esoc}}
 \and T.~Paumard\inst{\ref{lesia}}
 \and K.~Perraut\inst{\ref{ipag}}
 \and G.~Perrin\inst{\ref{lesia}}
 \and N.~Pourr\'e\inst{\ref{ipag}}
 \and L.~Pueyo\inst{\ref{stsci}}
 \and D.~C.~Ribeiro\inst{\ref{mpe}}
 \and E.~Rickman\inst{\ref{esa}}
 \and Z.~Rustamkulov\inst{\ref{jhueps}}
 \and J.~Shangguan\inst{\ref{peking},\ref{mpe}}
 \and T.~Shimizu \inst{\ref{mpe}}
 \and D.~Sing\inst{\ref{jhupa},\ref{jhueps}}
 \and J.~Stadler\inst{\ref{mpa},\ref{origins}}
 \and O.~Straub\inst{\ref{origins}}
 \and C.~Straubmeier\inst{\ref{cologne}}
 \and E.~Sturm\inst{\ref{mpe}}
 \and L.~J.~Tacconi\inst{\ref{mpe}}
 \and E.F.~van~Dishoeck\inst{\ref{leiden},\ref{mpe}}
 \and A.~Vigan\inst{\ref{lam}}
 \and F.~Vincent\inst{\ref{lesia}}
 \and S.~D.~von~Fellenberg\inst{\ref{bonn}}
 \and J.~J.~Wang\inst{\ref{northwestern}}
 \and F.~Widmann\inst{\ref{mpe}}
 \and J.~Woillez\inst{\ref{esog}}
 \and S.~Yazici\inst{\ref{mpe}}
 \and  the GRAVITY Collaboration
 }
\institute{
   European Southern Observatory, Karl-Schwarzschild-Stra\ss e 2, 85748 Garching, Germany\label{esog} \and
   LIRA, Observatoire de Paris, Universit\'e PSL, CNRS, Sorbonne Universit\'e, Universit\'e de Paris, 5 place Jules Janssen, 92195 Meudon, France\label{lesia}      \and
   Institute of Astronomy, University of Cambridge, Madingley Road, Cambridge CB3 0HA, United Kingdom\label{cam} \and
   Leiden Observatory, Leiden University, P.O. Box 9513, 2300 RA Leiden, The Netherlands\label{leiden} \and
   Department of Physics \& Astronomy, Johns Hopkins University, 3400 N. Charles Street, Baltimore, MD 21218, USA\label{jhupa}  \and
   Space Telescope Science Institute, 3700 San Martin Drive, Baltimore, MD 21218, USA\label{stsci} \and
   Division of Space Research \&\ Planetary Sciences, Physics Institute, University of Bern, Gesellschaftsstr.~6, 3012 Bern, Switzerland\label{bernWP} \and
   Max Planck Institute for Astronomy, K\"onigstuhl 17, 69117 Heidelberg, Germany\label{mpia} \and
   Fakult\"{a}t f\"{u}r Physik, Universit\"{a}t Duisburg-Essen, Lotharstraße 1, 47057 Duisburg, Germany\label{duisburg}
   Universidade de Lisboa - Faculdade de Ci\^encias, Campo Grande, 1749-016 Lisboa, Portugal\label{lisboa} \and
   CENTRA - Centro de Astrof\' isica e Gravita\c c\~ao, IST, Universidade de Lisboa, 1049-001 Lisboa, Portugal\label{centra} \and
   Univ. Grenoble Alpes, CNRS, IPAG, 38000 Grenoble, France\label{ipag} \and
   Center for Interdisciplinary Exploration and Research in Astrophysics (CIERA) and Department of Physics and Astronomy, Northwestern University, Evanston, IL 60208, USA\label{northwestern} \and
   Max Planck Institute for extraterrestrial Physics, Giessenbachstra\ss e~1, 85748 Garching, Germany\label{mpe} \and
   Aix Marseille Univ, CNRS, CNES, LAM, Marseille, France\label{lam} \and
   Université Côte d’Azur, Observatoire de la Côte d’Azur, CNRS, Laboratoire Lagrange, France\label{cotedazur} \and
  STAR Institute, Universit\'e de Li\`ege, All\'ee du Six Ao\^ut 19c, 4000 Li\`ege, Belgium\label{liege} \and
   Department of Astrophysical \& Planetary Sciences, JILA, Duane Physics Bldg., 2000 Colorado Ave, University of Colorado, Boulder, CO 80309, USA\label{boulder} \and
   1.\ Institute of Physics, University of Cologne, Z\"ulpicher Stra\ss e 77, 50937 Cologne, Germany\label{cologne} \and
   Max Planck Institute for Radio Astronomy, Auf dem H\"ugel 69, 53121 Bonn, Germany\label{bonn} \and
   Universidade do Porto, Faculdade de Engenharia, Rua Dr.~Roberto Frias, 4200-465 Porto, Portugal\label{porto} \and
   School of Physics, University College Dublin, Belfield, Dublin 4, Ireland\label{dublin} \and
   Astrophysics Group, Department of Physics \& Astronomy, University of Exeter, Stocker Road, Exeter, EX4 4QL, United Kingdom\label{exeterAstro} \and
   Departments of Physics and Astronomy, Le Conte Hall, University of California, Berkeley, CA 94720, USA\label{ucb} \and
   European Southern Observatory, Casilla 19001, Santiago 19, Chile\label{esoc} \and
   Advanced Concepts Team, European Space Agency, TEC-SF, ESTEC, Keplerlaan 1, NL-2201, AZ Noordwijk, The Netherlands\label{actesa} \and
   University of Exeter, Physics Building, Stocker Road, Exeter EX4 4QL, United Kingdom\label{exeter} \and
   French-Chilean Laboratory for Astronomy, IRL 3386, CNRS and U. de Chile, Casilla 36-D, Santiago, Chile\label{french_chilean_lab} \and
   Center for Space and Habitability, University of Bern, Gesellschaftsstr.~6, 3012 Bern, Switzerland\label{bernCSH} \and
   Academia Sinica, Institute of Astronomy and Astrophysics, 11F Astronomy-Mathematics Building, NTU/AS campus, No. 1, Section 4, Roosevelt Rd., Taipei 10617, Taiwan\label{sinica} \and
   European Space Agency (ESA), ESA Office, Space Telescope Science Institute, 3700 San Martin Drive, Baltimore, MD 21218, USA\label{esa} \and
   Department of Earth \& Planetary Sciences, Johns Hopkins University, Baltimore, MD, USA\label{jhueps} \and
   Kavli Institute for Astronomy and Astrophysics, Peking University, Beijing 100871, People's Republic of China\label{peking} \and
   Max Planck Institute for Astrophysics, Karl-Schwarzschild-Str. 1, 85741 Garching, Germany\label{mpa} \and
   Excellence Cluster ORIGINS, Boltzmannstraße 2, D-85748 Garching bei München, Germany\label{origins}
}

\date{Received XXX; accepted YYY}


\abstract
{
Inferring the likely formation channel of giant exoplanets and brown dwarf companions from orbital and atmospheric observables remains a formidable challenge.
Further and more precise directly measured dynamical masses of these companions are required to inform and gauge formation, evolutionary, and atmospheric models.
We present an updated study of the recently discovered companion to HIP~99770 based on observations conducted with the near-infrared interferometer VLTI/GRAVITY.
}
{
Through renewed orbital and spectral analyses based on the GRAVITY data, we characterise HIP~99770~b to better constrain its orbit, dynamical mass, and atmospheric properties, as well as to shed light on its likely formation channel.
}
{
Upon inclusion of the new high-precision astrometry epoch, we ran an orbit fit to further constrain the dynamical mass of the companion and the orbit solution.
We also analysed the GRAVITY K-band spectrum, placing it into context with literature data, and extracting magnitude, age, spectral type, bulk properties and atmospheric characteristics of HIP~99770~b.
}
{
We detected the companion at a radial separation of \SI{417}{mas} from its host. The new orbit fit yields a dynamical mass of $17_{-5}^{+6}\,\mathrm{M}_\mathrm{Jup}$ and an eccentricity of $0.31_{-0.12}^{+0.06}$. We also find that additional relative astrometry epochs in the future will not enable further constraints on the dynamical mass due to the dominating relative uncertainty on the Hipparcos--\textit{Gaia} proper motion anomaly that is used in the orbit-fitting routine. The publication of Gaia DR4 will likely ease this predicament.
Based on the spectral analysis, we find that the companion is consistent with spectral type L8 and exhibits a potential metal enrichment in its atmosphere.
Adopting the AMES-DUSTY model to infer its age, within its dynamical mass constraint the companion conceivably corresponds to either a younger ($28_{-14}^{+15}\,\mathrm{Myr}$) object with a mass just below the deuterium-burning limit or an older ($119_{-10}^{+37}\,\mathrm{Myr}$) body with a mass just above the deuterium-burning limit.
}
{
These results do not yet allow for a definite inference of the companion's formation channel.
Nevertheless, the new constraints on its bulk properties and the additional GRAVITY spectrum presented here will aid future efforts to determine the formation history of HIP~99770~b.
}

\keywords{Astrometry --
         Techniques: interferometric, spectroscopic --
         Planets and satellites: gaseous planets, brown dwarfs, formation --
         }

\maketitle


\section{Introduction}
\label{section_introduction}

Gas giant exoplanets that are accessible to direct-imaging studies are rare. Unbiased blind surveys have provided only a few detections (e.g.\,\citealt{nielsen_gpi_demo, vigan_shine_demo, chomez_shine_v}; for a review, see \citealt{bowler2016imaging}) and demonstrated that targeted studies, for instance informed by clues stemming from other detection techniques such as astrometric monitoring of potential host stars, are the only way to efficiently build a population-level sample of directly imaged exoplanets.

The source HIP~99770 was identified as exhibiting a significant proper motion anomaly (PMa; see \citealt{kervella_stellar_and_substellar_2022, brandt_hipparcos_gaia}), the discrepancy between the proper motion of a star as observed by the Hipparcos \citep{schuyer_hipparcos} and \textit{Gaia} missions \citep{gaia_mission}.
Such a PMa can be caused by an orbiting companion inducing reflex motion in the star that manifests itself in its projected movement on the sky plane, and thus, in its proper motion.
A dedicated direct-imaging follow-up using SCExAO/CHARIS \citep{jovanovic_SCExAO, groff_charis} on the Subaru Telescope and Keck/NIRC2 revealed a substellar companion with a dynamical mass of $16.1_{-5.0}^{+5.4}\,\mathrm{M}_\mathrm{Jup}$ \citep{currie_hip99770b}.
Hence, HIP~99770 was the first system to yield a detection of a companion identified as a planet after it was targeted specifically on the basis of tentative astrometric evidence that indicated the existence of an orbiting object.
The loose mass constraint places the companion in close proximity to the deuterium-burning threshold, implying a need for further observations to pin down its mass\footnote{While the deuterium-burning threshold, which is traditionally assumed to be located at \SI{13}{M_{Jup}} (e.g.\ \citealt{burrows_bd_theory, spiegel_d_burning,molliere_d_burning}), is the conventional choice of classifier between planets and BDs, others have been proposed. For instance, there is evidence that suggests that the turn-over mass of the companion mass function is located far beyond \SI{13}{M_{Jup}} in the region between \SI{25}{} and \SI{40}{M_{Jup}} (e.g.\,\citealt{sahlmann2011search, ma2014statistical, reggiani2016vltnaco, kiefer2019detection, stevenson2023combing})}.

The necessity of follow-up studies of this object becomes all the more apparent when we consider the relative scarcity of what are loosely referred to as super-Jupiters \citep{carson_super_jupiter}, that is, substellar companions with masses at about the deuterium-burning threshold.
These objects tend to defy the commonly proposed planet formation mechanisms. On the one hand, the disc-instability pathway, which posits that planets form via the gravitational collapse of the circumstellar disc \citep{cameron_grav_inst, boss_grav_inst, kratter_grav_inst}, is predicted to form more massive companions well within the brown dwarf or even stellar mass regime \citep{forgan_rice_towards_pop_synth}.
On the other hand, formation via core accretion in the outer disc is difficult to justify because the required timescales at these distances from the host are significantly longer than the typical disc-dissipation timescale \citep{lambrechs_rapid_growth}.
The different pathways are expected to leave telltale imprints on the orbital geometry of the system implying that precisely constraining the orbital elements of a given companion can reveal clues as to its formation history (e.g.\ \citealt{bitsch_eccentricity_distr,marleau_hip65426b}).
Given its long period orbit (approximately \SI{50}{years}; \citealt{currie_hip99770b}) and its recent discovery based on data collected between 2020 and 2021, the current orbit coverage of HIP~99770~b is limited. An extension of the available astrometric baseline will help us improve the orbital solution.

An alternative and complementary route towards probing the formation history of a given companion is the characterisation of its atmosphere. Observables such as the atmospheric metallicity or elemental abundance ratios, which are accessible via a spectroscopic analysis of the flux emitted by the companion, can add to the conclusions drawn from orbital considerations (e.g.\ \citealt{molliere_interpreting_atmo_composition}).
For the specific case of HIP~99770~b, a recent high-resolution spectroscopic follow-up yielded constraints on the metallicity and elemental abundance ratios of the companion \citep{zhang_hip99770b}. These were unable to exclude either formation pathway, however.
In addition to these continuum-removed high-resolution K-band spectra, a new data set that preserves the continuum emission component of the companion (albeit at a lower resolution) can shed new light on the atmospheric properties of the companion.

Here, we present new VLTI/GRAVITY \citep{GRAVITY_Collaboration_First_light} observations of HIP~99770~b, describe how they facilitate constraints on the orbital geometry of the system and act as yet another window into the companion atmosphere.
GRAVITY is a near-infrared interferometric instrument at the European Southern Observatory's Very Large Telescope (VLT).
Previous studies using the instrument for exoplanet observations have demonstrated an astrometric accuracy of \SI{50}{\micro as} \citep{lacour_first_detection}.

This paper is arranged as follows: Section~\ref{section_observations} gives an overview of the observational data we used. The orbital and spectral analyses of the obtained data are presented and discussed in Sect.~\ref{section_orbital_analysis} and \ref{section_spectral_analysis}, respectively. The conclusions of our study are laid out in Sect.~\ref{section_conclusion}.


\section{Observations and data reduction}
\label{section_observations}

\subsection{Previous observations}
\label{subsect_previous_observations}
This work draws upon data that were collected by previous studies of the HIP~99770 system. We used the astrometric epochs and the photometric and spectral data obtained by \citet{currie_hip99770b}.
The CHARIS spectrum contained therein consists of \SI{22}{} channels covering the wavelength range between \SI{1.16}{} and \SI{2.37}{\micro m} at a resolution of approximately \SI{20}{}.
Additionally, the orbital fit performed in Sect.~\ref{section_orbital_analysis} is partially based on the host star absolute astrometry as listed in the Hipparcos-\textit{Gaia} Catalogue of Accelerations (HGCA; \citealt{brandt_hipparcos_gaia}).

\subsection{VLTI/GRAVITY}
\label{subsect_vlti_observations}
We observed HIP~99770~b with the GRAVITY instrument on 31 May and 2 July 2023. These observations were obtained on technical time and in the framework of the ExoGRAVITY Large programme (ESO ID 1104.C-0651 \citealt{lacour_exogravity_lp}). An observation log for both epochs can be found in Table \ref{table_obs_log}. Notably, they were taken using different observing modes. For the first epoch, we used the dual-field on-axis mode, while the second epoch was obtained in dual-field off-axis mode.
Importantly, the latter mode enables a higher signal-to-noise ratio (S/N) because the entire light of the sources that is collected by the four unit telescopes is injected into the fringe tracker and science fibre channels by means of a rooftop mirror. This is not the case for on-axis observations, where a beam splitter directs only half the light into the different channels. A comprehensive comparison of the two observing modes is given by \citet{nowak_catalogue_of_dual_field}.
For both observations, the placement of the science fibre was informed by orbital fits applied to the available direct detections of the companion (see Sect.~\ref{subsect_previous_observations} and Table~\ref{table_astrometry}).

\begin{figure}
        \centering
        \includegraphics[width=0.9\columnwidth]{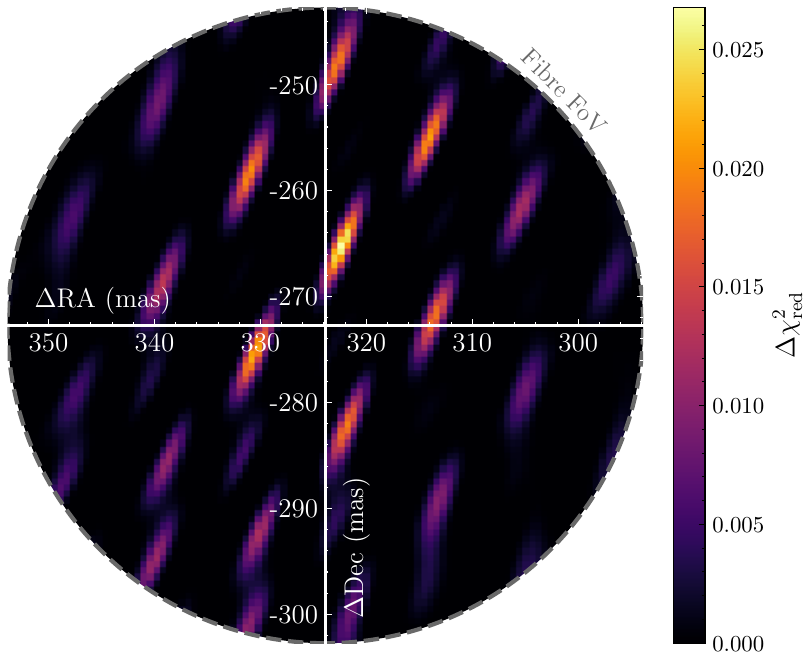}––
        \caption{Detection of HIP~99770~b on 31 May 2023. The circular panel indicates the field of view of the GRAVITY science fibre. The map within visualises the $\Delta \chi_\mathrm{red}^2$ over a two-dimensional grid of positions relative to the host star, which is located at the origin of the axes, and thus, outside the field of view. Each $\Delta \chi_\mathrm{red}^2$ value is the difference between how well a model with and without a coherent point source of light at a given position $(\Delta \mathrm{RA}, \Delta \mathrm{Dec})$ describes the interferometric observables. Thus, the strongest peak within the field of view corresponds to the position of the companion, while the secondary peaks (the so-called side lobes) are typical for interferometric observations and mostly depend on the u-v plane coverage.}
        \label{figure_detection}
\end{figure}

We used the ESO GRAVITY instrument pipeline 1.6.4b1 for the data reduction. This yielded the so-called ``\texttt{astroreduced}'' byproduct, from which the relative astrometry between host and companion can be extracted following the standard exoplanet dual-field data processing outlined in Appendix A of \citet{nowak_peering_into}.
The first detection on 31 May 2023 is visualised in Fig.~\ref{figure_detection}, the resulting relative astrometry epoch is listed in Table \ref{table_astrometry}.
Because the off-axis observation mode was employed for the second epoch, the collected data necessitated a special reduction procedure. To measure the metrology zero point, we performed a swap observation of a calibration binary, which in this case was HD~196885~AB. The binary observation was based on a poor relative position prediction, and the pointing of the single-mode fibres was therefore off by approximately \SI{35}{mas}. The field aberrations resulting from this pointing deviation \citep{2021A&A...647A..59G} caused a bias in the zero-point estimate of up to 30 degrees. As a consequence, the astrometric measurement of the science target HIP~99770~b was uncertain, which caused us to discard this astrometric epoch from our orbital analysis.

For each epoch, we also obtained a K-band companion-to-host contrast spectrum ranging from \SI{1.97}{} to \SI{2.48}{\micro m} at a resolution of $R=\SI{500}{}$. Depending on the angular distance between the astrometric position of the companion and the centre of the science fibre, we needed to correct the spectra for throughput losses. A detailed description of how this is accounted for is provided in Appendix~\ref{app_section_correcting_for_tp_losses}.
To convert the companion contrast spectra measured by GRAVITY into companion flux spectra that could be used for the atmospheric analysis and modelling, we first multiplied them with the flux spectrum of the amplitude reference source, that is, the host star HIP~99770~A for the on-axis, and the binary (swap) calibrator HD~196885~AB for the off-axis epoch.
To obtain model spectra of these targets, we fitted a \texttt{BT-NextGen} and a \texttt{BT-Settl (CIFIST)} solar metallicity stellar model atmosphere \citep{allard_homeier_freytag_models} to archival photometric data from \textit{Gaia} and Tycho using the \texttt{species} toolkit \citep{stolker_species}. Additionally, we included the \textit{Gaia} XP spectrum of HIP~99770~A and HD~196885~AB in these fits. When we also incorporated 2MASS photometry, the inferred stellar model parameters did not change by more than $\SI{1}{\sigma}$.
Based on \texttt{PyMultiNest} \citep{feroz_multimodal_nested_sampling, feroz_multinest, buchner_x_ray_spectral}, the nested-sampling approach implemented within \texttt{species} enables the inference of the posterior distributions of the individual model parameters. Following \citet{cardelli1989}, we also accounted for extinction by the interstellar medium that might affect HIP~99770~A by setting the V-band extinction parameter, $A_V$, to \SI{0.043}{mag} \citep{murphy2017}.
To further guide the spectral fits to physically plausible solutions, we invoked Gaussian priors on the mass of HIP~99770~A of $M = 1.8\pm0.2~\rm{M}_\odot$ \citep{currie_hip99770b} and on the effective temperatures and masses of the two binary components of HD~196885~AB using the values and uncertainties reported in Table~1 of \citet{chauvin2023} ($T_\text{eff,A} = 6340\pm19$~K, $T_\text{eff,B} = 3660\pm190$~K, $M_\text{A} = 1.3\pm0.1~\rm{M}_\odot$, $M_\text{B} = 0.45\pm0.01~\rm{M}_\odot$).
We converted the M1$\pm$1 spectral type constraint of component B that is reported in this table into an effective temperature constraint using Table~5 of \citet{pecaut2013}. For more details on the conversion of contrast to flux, we refer to Appendix \ref{app_section_GRAVITY_spectral_reduction}. The mass and effective temperature values of HIP~99770~A inferred from our fit are within $2\sigma$ of the value reported by \citet{murphy2017}, those of HD~196885~AB are within $3\sigma$ of the values reported by \citet{chauvin2023}.

The resulting on- and off-axis flux spectra normalised by the respective amplitude references are significantly offset relative to each other.
Since the off-axis reference (swap) binary HD~196885~AB was observed with a significant mispointing of the fibre, a lower throughput is expected (estimated at $\approx45$\% in Table~\ref{table_obs_log}). This biased our photometric calibration. The on-axis observations of HIP~99770~b, on the other hand, were calibrated quasi-simultaneously as a result of the continuous monitoring of the host star spectrum, with GRAVITY science spectrometer observations interspersed between the planet observations. The on-axis spectrum is therefore more reliable, and we opted for scaling the off-axis companion flux spectrum such that it matched the on-axis spectrum best. The best-fitting scale factor was found to be \SI{61}{\percent}.

\begin{figure}
        \centering
        \includegraphics[width=0.98\columnwidth]{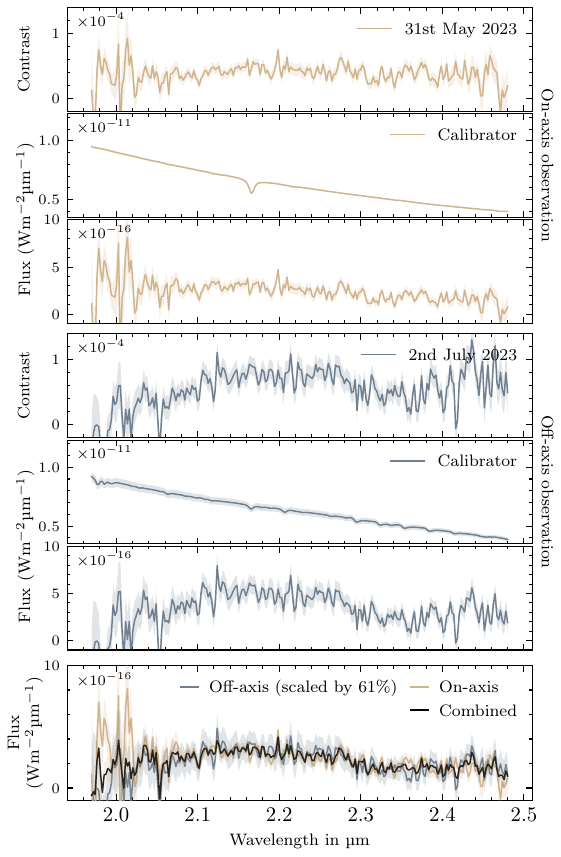}
        \caption{Ingredients and outcome of the spectral data reduction.
        The top three panels show the GRAVITY on-axis contrast spectrum, the associated stellar calibration flux spectrum, and the resulting companion flux spectrum.
        The same is visualised for the off-axis observation in the three middle panels.
        Finally, the bottom panel shows the on- and off-axis companion flux spectra alongside the combination of the two. This K-band spectrum of HIP~99770~b is used in Sect.~\ref{section_spectral_analysis}. \\}
        \label{figure_spectral_data_reduction}
\end{figure}

Finally, we combined the companion flux spectra for the two epochs into a single covariance-weighted mean spectrum that we used for the analysis presented in Sect.~\ref{section_spectral_analysis}. The spectral reduction process is visualised in Fig.~\ref{figure_spectral_data_reduction}.
This is the first time a GRAVITY exoplanet off-axis spectrum without an on-axis amplitude reference is published.

\begin{table*}[t]
    \centering
    \caption{GRAVITY observation log of the target HIP~99770~b and of the swap calibrator HD\,196885\,AB.}
    \label{table_obs_log}
    \resizebox{0.95\textwidth}{!}{%
    \begin{tabular}{cccccccccc}
    \toprule
    Target & Start & End & NEXP/NDIT/DIT(s) & Airmass & $\tau_0$ & Seeing & \multicolumn{2}{c}{Fibre placement} & $\gamma$ \\ 
     & (UTC) & (UTC) &  &  & (ms) & (arcsec) & $\Delta$RA (mas) & $\Delta$Dec (mas) & \\
    \midrule
    \midrule
     \multicolumn{10}{c}{31st May} \\
    \midrule
    \object{HIP 99770} A & 09:19:59 & 09:25:04 & 2/96/0.5 & 2.19 & 8.2 & 0.62 & 0.0 & 0.0 & \SI{100}{\percent} \\
    \object{HIP 99770 b} & 09:27:53 & 09:39:12 & 2/32/10 & 2.25 & 7.2 & 0.63 & 323.87 & -272.72 & \SI{96.3}{\percent} \\
    \midrule
    \multicolumn{10}{c}{2nd July 2023} \\
    \midrule 
    HIP 99770 b & 07:05:36 & 07:32:34 & 3/12/30 & 2.22 & 2.9 & 0.55 & 323.6 & -261.6 & \SI{99.8}{\percent} \\
    \object{HD 196885} AB & 07:43:03 & 07:56:17 & 4/32/3 & 1.31 & 2.4 & 0.91 & 285.0 & 430.0 & \SI{44.8}{\percent} \\
    \bottomrule
    \end{tabular}
    }
    \tablefoot{The reported start and end times indicate the duration of the full observation, consisting of a series of NEXP exposures, each subdivided into NDIT detector integrations per exposure, that were taken over a duration of DIT. $\tau_0$ is the mean atmospheric coherence time during the observation, and the fibre placement columns display where the centre of the science fibre was located relative to the host star at the time of observation. The positions are given as on-sky Cartesian coordinates in separation relative to the host star along the right ascension axis, $\Delta$RA, and the declination axis, $\Delta$Dec. The last column lists the normalised coupling efficiency, $\gamma$, by which the respective contrast spectrum needs to be divided to correct for the throughput loss resulting from an imperfect alignment of the fibre centre and the companion (see Appendix~\ref{app_section_correcting_for_tp_losses}). \\
    }
\end{table*}

\begin{table}[t]
    \centering
    \caption{Astrometric detections of HIP~99770~b.}
    \label{table_astrometry}
    \resizebox{\columnwidth}{!}{%
    \begin{tabular}{cccccccc}
    \toprule
    MJD & Instrument & $\Delta$RA & $\Delta$Dec & $\rho$ \\
    (days) & & (mas) & (mas) & \\
    \midrule
    \midrule
    59059 & CHARIS & \SI{263 (4)}{} & \SI{-367 (5)}{} & --- \\
    59093 & CHARIS & \SI{263 (5)}{} & \SI{-366 (5)}{} & --- \\
    59353 & CHARIS & \SI{280 (4)}{} & \SI{-343 (4)}{} & --- \\
    59368 & NIRC2 & \SI{286 (6)}{} & \SI{-337 (6)}{} & --- \\
    59408 & CHARIS & \SI{286 (4)}{} & \SI{-338 (4)}{} & --- \\
    59504 & CHARIS & \SI{292 (4)}{} & \SI{-327 (4)}{} & --- \\
    \midrule
    60095.40 & GRAVITY & \SI{322.27 (10)}{} & \SI{-265.09 (18)}{} & \SI{-0.88}{} \\
    60127.31\hyperlink{astrometry_corrupted}{\textsuperscript{*}} & GRAVITY & \SI{322.21 (36)}{} & \SI{-260.66 (49)}{} & \SI{-0.88}{} \\
    \bottomrule
    \end{tabular}
    }
    \tablefoot{The astrometric position of the companion relative to the host star at each epoch is presented as separation components along the right ascension axis, $\Delta$RA, and the declination axis, $\Delta$Dec. The CHARIS and NIRC2 epochs were originally published by \citet{currie_hip99770b}. For these, the time of observation is rounded to the nearest full day. The GRAVITY data we present also comprise a correlation coefficient, $\rho$, between $\Delta$RA and $\Delta$Dec. \\
    \hypertarget{astrometry_corrupted}{\textsuperscript{*}} Since this astrometric epoch is corrupted, it was not used anywhere in this study.
    }
\end{table}


\section{Orbital analysis}
\label{section_orbital_analysis}
\begin{figure}
        \centering
        \includegraphics[width=0.99\columnwidth]{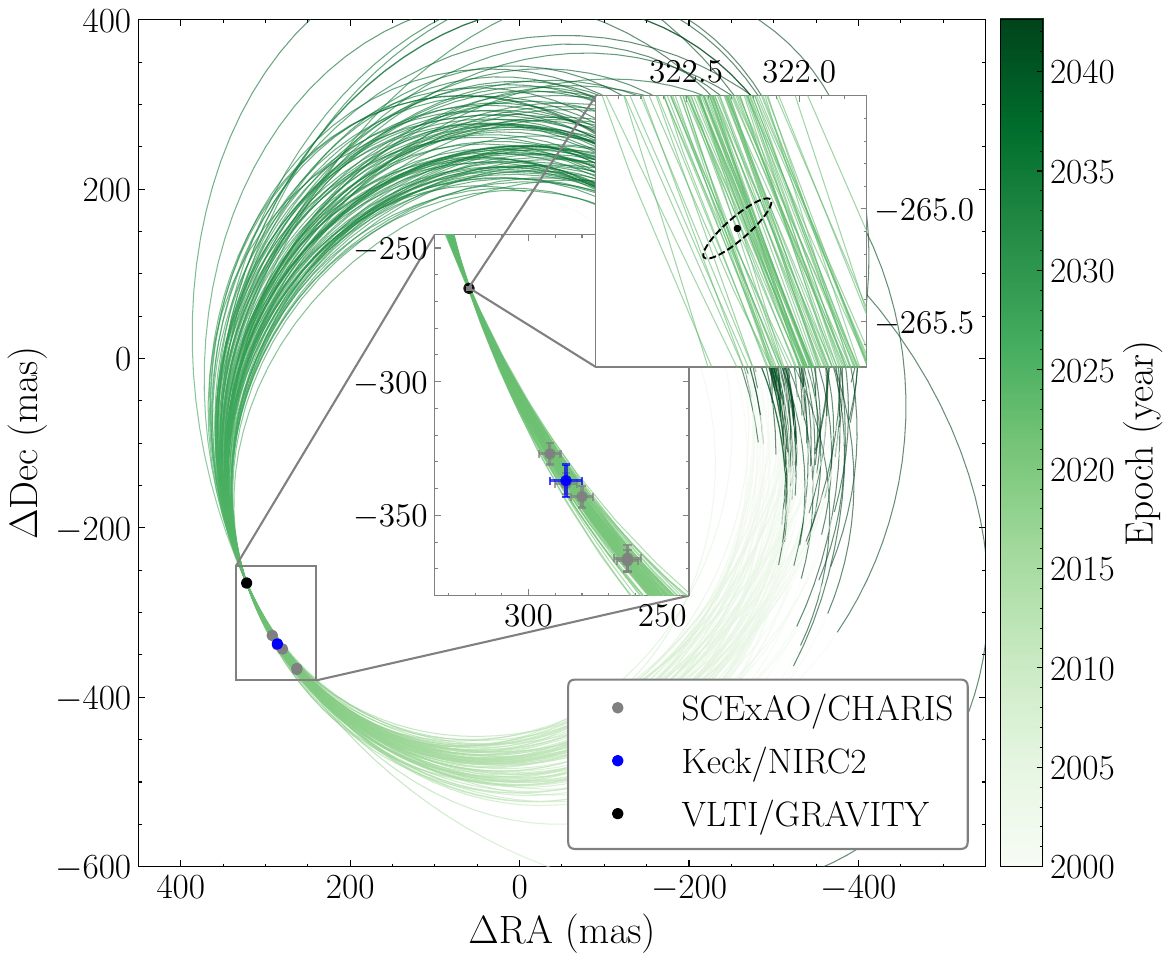}
        \caption{Orbit of HIP~99770~b relative to its host, i.e., fixed at the origin. The visualised orbits were generated from a random subset of \SI{100}{} parameter sets drawn from the posterior sampling. While the previously available astrometric epochs obtained by SCExAO/CHARIS and Keck/NIRC2 are shown along with their respective uncertainties in $\Delta$RA and $\Delta$Dec, knowledge of the correlation coefficient, $\rho$, allowed us to visualise the VLTI/GRAVITY epoch at a \SI{1}{\sigma} confidence interval.}
        \label{figure_orbit}
\end{figure}

\begin{figure}
        \centering
        \includegraphics[width=0.9\columnwidth]{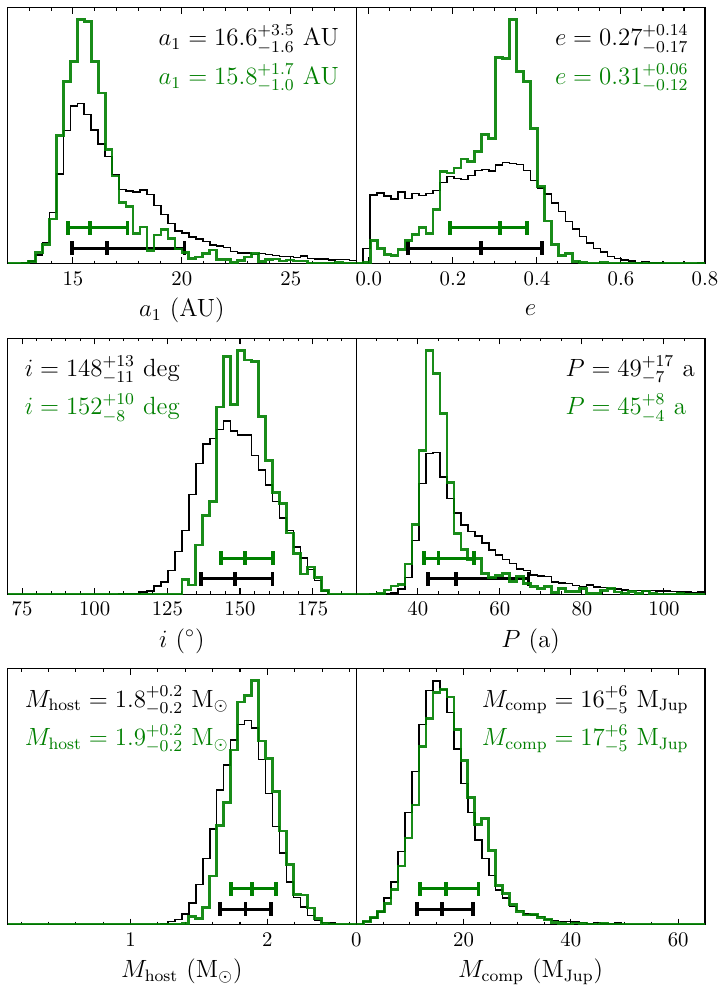}
        \caption{Marginalised posterior distributions of a subset of the fitted orbital parameters. The black posteriors were sampled considering the previously available data only, while the green posteriors result from the inclusion of the new GRAVITY epoch from 31 May 2023. The median values and intervals between the 16th and 84th percentiles of the distribution are indicated by the horizontal bars. The inferred values resulting from the two runs, which were rounded to the respective significant figure, are displayed at the top of each panel. While technically not one of the parameters explored by the walkers during the sampling procedure, we also plot the period, $P$, as computed from the other posteriors via Kepler's third law. The full posterior samplings can be found in Appendix~\ref{app_section_additional_orbital_fitting_plots}.
        }
        \label{figure_posteriors}
\end{figure}

To obtain constraints on the companion mass and orbital parameters, we performed an orbit fit of the relative astrometry epochs presented in Table~\ref{table_astrometry} using the \texttt{orbitize!} package \citep{blunt_orbitize}. Its MCMC sampling routine allowed us to include the absolute astrometry of the host star collected by the Hipparcos and \textit{Gaia} missions (conveniently presented by \citealt{brandt_hipparcos_gaia}).
The \texttt{orbitize!} sampling procedure is an implementation of \texttt{ptemcee} \citep{vousden2016dynamic}, a parallel-tempered affine-invariant Markov chain Monte Carlo (MCMC) algorithm based on \texttt{emcee} \citep{foreman2013emcee}.
It was carried out using \SI{20}{} temperatures and \SI{100}{} walkers taking \SI{4e4}{} burn-in steps and another \SI{4e4}{} actual sampling steps each.
These walkers were let loose on nine free parameters, namely the semi-major axis, $a$, the eccentricity, $e$, the inclination, $i$, the argument of periastron, $\omega$, the longitude of the ascending node, $\Omega$, the relative epoch of periastron\footnote{Using the relative epoch of periastron, $\tau$, is a sleight of hand that simplifies setting the prior range boundaries. In combination with a reference epoch ($t_\mathrm{ref} = \SI{58849}{MJD}$) and the orbital period, $P$, it relates to the actual time of periastron passage, $t_\mathrm{p}$, via $\tau=(t_\mathrm{p}-t_\mathrm{ref})/P$ and is thus a dimensionless quantity.\label{footnote_tau}}, $\tau$, the parallax, $\pi$, the stellar mass, $M_\mathrm{host}$, and the companion mass, $M_\mathrm{comp}$.
The priors we chose for the individual free parameters are listed in Table~\ref{table_orbital_priors_and_posteriors}. 
We carried out two separate sampling runs. The first run was performed using the Hipparcos-\textit{Gaia} stellar astrometry and the previous astrometric companion detections \citep{currie_hip99770b} only. We then resampled the posterior distributions upon inclusion of our own GRAVITY detection from 31 May 2023.
With the two resulting posterior samplings in hand we investigated how the addition of a GRAVITY detection changed the inferred parameter values and their uncertainties.
By means of visual inspection of the chain traces for the individual parameters, we confirmed that the walker chains converged for all parameters and for both runs.
The two posterior samplings of the orbital parameters are shown in Fig.~\ref{figure_posteriors}. A random subset of the posterior sampling was used to visualise the projection of the companion orbit onto the sky plane in Fig.~\ref{figure_orbit}.

The bottom right panel of Fig.~\ref{figure_posteriors} shows that the resampling procedure resulted in a slightly increased companion mass of $16.7_{-4.8}^{+6.0}\,\mathrm{M}_\mathrm{Jup}$\footnote{\label{non_rounded_values}These values have not been rounded to their respective significant figures to facilitate better comparison.}.

To investigate the extent to which this result is driven by the uniform companion mass prior, we ran an additional sampling procedure using a log-uniform prior distribution instead. This yielded a lower posterior mass of $13.5_{-4.4}^{+5.2}\,\mathrm{M}_\mathrm{Jup}$\footref{non_rounded_values}, a behaviour consistent with the findings by \citet{currie_hip99770b}.
Along the same lines, we examined the dependence of the inferred companion mass on the stellar mass prior. Retaining the standard deviation of \SI{0.2}{M_\odot}, we performed further posterior samplings based on underlying Gaussian priors that were shifted from the original \SI{1.8}{} to \SI{1.7}{} in the first and \SI{1.9}{M_\odot} in the second run.
These modifications resulted in a companion mass of $14.7_{-4.4}^{+5.0}\,\mathrm{M}_\mathrm{Jup}$\footref{non_rounded_values} (decrease of \SI{12}{\percent}) and $16.8_{-5.3}^{+5.7}\,\mathrm{M}_\mathrm{Jup}$\footref{non_rounded_values} (increase of \SI{0.5}{\percent}), respectively. A comparison between the resulting posterior mass distributions is shown in the top right panel of Fig.~\ref{figure_full_orbit_corner_plot}.
Since these adjustments to the shapes and positions of the priors affect larger posterior displacements than the inclusion of the GRAVITY astrometry epoch, the slight shift in companion mass evident in Fig.~\ref{figure_posteriors} does not necessarily amount to a robust inference of a higher dynamical mass.

The manifest inability to further constrain the companion mass comes as somewhat of a surprise given the high angular resolution of the new GRAVITY astrometry epoch. It is, however, not driven by the individual astrometric epochs or the amount thereof, but by the large relative uncertainty of the \textit{Gaia}-Hipparcos PMa. Following \citet{kervella_stellar_and_substellar_2019}, we computed the PMa as the difference between the \textit{Gaia} DR3 proper motion vector, $\mu_\mathrm{G3}$, and the long-term Hipparcos-\textit{Gaia} DR3 proper motion vector, $\mu_\mathrm{HG3}$, which both consist of a right ascension and declination component. Carrying the respective uncertainties, we found the PMa to be
\begin{align*}
    (\vec{\mu}_\mathrm{G3} - \vec{\mu}_\mathrm{HG3}) &=
    \begin{pmatrix}
    68.09 \pm 0.12 \\
    69.40 \pm 0.14 
    \end{pmatrix}
    -
    \begin{pmatrix}
    68.236 \pm 0.012 \\
    69.671 \pm 0.012 
    \end{pmatrix}
    \\
    &= 
    \begin{pmatrix}
    0.15 \pm 0.12 \\
    0.27 \pm 0.14 
    \end{pmatrix}
    ,
\end{align*}
where all components are given in units \SI{}{mas \per yr}, and the proper motion values were taken from \citet{brandt_hipparcos_gaia}. Thus, the relative uncertainties on the RA and Dec components are approximately \SI{80}{} and \SI{51}{\percent}, respectively. This large PMa uncertainty dominates so thoroughly that even additional GRAVITY epochs in the future would not help constrain the companion mass further. Tighter error bars on the mass can therefore only be obtained once a more precise PMa will be available in \textit{Gaia} DR4. To substantiate this finding, we performed several orbital resampling runs including varying numbers of mock GRAVITY epochs predicted from the current best fit. While these additional artificial epochs served to narrow down the posterior distributions of the remaining orbital parameters, they likewise fail to further constrain the companion mass. For instance, when we iteratively generated an additional three GRAVITY astrometric mock epochs based on the newly obtained orbit solution, we found an improvement of approximately \SI{1}{\percent} in relative uncertainty of the companion semi-major axis, but no such gain is achieved in the precision of its mass.

While the addition of the GRAVITY astrometry epoch yielded no improved constraints on the companion mass, it did pin down the orbital semi-major axis at $15.8_{-1.0}^{+1.7}\,\mathrm{AU}$ and the eccentricity at $0.31_{-0.12}^{+0.06}$.
The strong constraint on the orbital eccentricity is a marked improvement on the previously available solution. While a circular orbit was conceivable prior to the GRAVITY detection \citep{currie_hip99770b}, this configuration is now decisively ruled out. HIP~99770~b therefore joins a growing number of substellar companions exhibiting elevated eccentricities. Examples of other recently detected eccentric companions are 51~Eri~b \citep{macintosh2015discovery} with $e=0.57_{-0.06}^{+0.08}$ \citep{dupuy2022limits} and Eps~Ind~A~b \citep{feng2019detection} with $e=0.40_{-0.18}^{+0.15}$ \citep{matthews2024temperate}.

\begin{table}[t]
    \centering
    \caption{Orbital parameter priors and posteriors.}
    \label{table_orbital_priors_and_posteriors}
    \renewcommand{\arraystretch}{1.4}
    \resizebox{\columnwidth}{!}{%
    \begin{tabular}{lcc}
    \toprule
    Parameter & Prior type and range & Posterior \\
    \midrule
    \midrule
    Semi-major axis (AU) & Log uniform [\SI{e-5}{}, \SI{e3}{}] & $15.8_{-1.0}^{+1.7}$ \\
    Eccentricity & Uniform [0, 1) & $0.31_{-0.12}^{+0.06}$ \\
    Inclination (deg) & Sine [0, $\pi$) & $152_{-8}^{+10}$ \\
    Argument of periastron (deg) & Uniform [0, $2\pi$) & $210_{-50}^{+90}$ \\
    Long. of ascending node (deg) & Uniform [0, $2\pi$) & $190_{-30}^{+120}$ \\
    Relative epoch of periastron\footref{footnote_tau} & Uniform [0, 1) & $0.36_{-0.06}^{+0.06}$ \\
    Parallax (mas) & Gaussian ($24.55 \pm 0.09$)\hyperlink{ref_parallax}{\textsuperscript{*}} & $24.56_{-0.08}^{+0.08}$ \\
    Stellar mass ($\mathrm{M}_{\odot}$) & Gaussian ($1.8 \pm 0.2$)\hyperlink{ref_stellar_mass}{\textsuperscript{\dag}} & $1.9_{-0.2}^{+0.2}$ \\
    Companion mass ($\mathrm{M}_\mathrm{Jup}$) & Uniform [1, 50] & $17_{-5}^{+6}$ \\
    \bottomrule
    \end{tabular}
    }
    \tablefoot{For the log uniform and uniform distributions, the square and rounded brackets denote inclusive and exclusive boundaries, respectively. The reported values were rounded to their significant figures. They correspond to the marginalised posterior distribution median and its differences to the 84th and 16th percentiles in superscript and subscript, respectively (see Fig.~\ref{figure_posteriors} and Appendix~\ref{app_section_additional_orbital_fitting_plots} for the marginalised posteriors). \\
    \hypertarget{ref_parallax}{\textsuperscript{*}} \citet{gaia_dr3} \\
    \hypertarget{ref_stellar_mass}{\textsuperscript{\dag}} \citet{currie_hip99770b}
    }
\end{table}

\section{Spectral analysis}
\label{section_spectral_analysis}

\subsection{Spectral classification}
\label{subsection_spectral_classification}
\begin{figure}
        \centering
        \includegraphics[width=0.9\columnwidth]{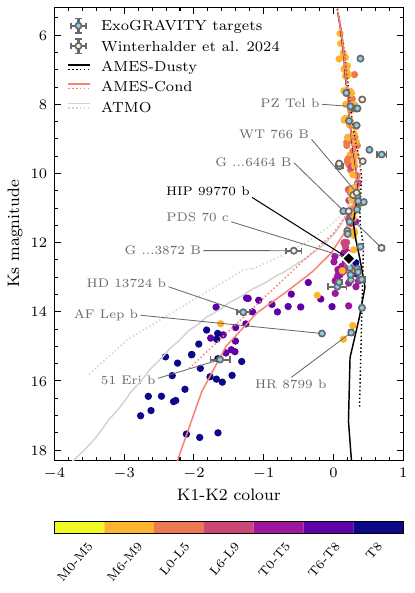}
        \caption{Colour-magnitude diagram showing HIP~99770~b, indicated by the black diamond, in relation to a literature population of low-mass stars, brown dwarfs, and exoplanets (see Appendix C of \citealt{bonnefoy_gj504} and references therein). All ExoGRAVITY targets (ESO ID 1104.C-0651; \citealt{lacour_exogravity_lp}) and the companions so far detected via the \textit{Gaia}--GRAVITY synergy \citep{winterhalder_gaia_gravity} are shown in light blue and ochre, respectively.
        Additionally, isochrones computed using the AMES-Dusty \citep{chabrier2000evolutionary, allard2001limiting}, AMES-Cond \citep{allard2001limiting, baraffe2003evolutionary}, and \texttt{ATMO} \citep{phillips_atmo} evolutionary models are shown for two ages: \SI{1}{Gyr} (solid lines), and \SI{100}{Myr} (dotted lines).
        The filters we used to extract these magnitudes and colours from the spectra are \texttt{Paranal}\texttt{/}\texttt{SPHERE}\texttt{.}\texttt{IRDIS\_B\_Ks}, \texttt{IRDIS\_D\_K12\_1}, and \texttt{IRDIS\_D\_K12\_2}.
        }
        \label{figure_cmd}
\end{figure}

We folded the GRAVITY K-band spectrum with the \texttt{Paranal}\texttt{/}\texttt{SPHERE}\texttt{.}\texttt{IRDIS\_B\_Ks}, 
\texttt{IRDIS\_D\_K12\_1}, and \texttt{IRDIS\_D\_K12\_2} filter profiles to place HIP~99770~b in a colour-magnitude diagram (CMD) and thereby place it in the context of a literature population of companion and free-floating brown dwarfs and exoplanets.
This folding procedure yielded absolute companion magnitudes of $M_\texttt{Ks} = \left(\SI{12.460 (14)}{}\right)\,\SI{}{mag}$, $M_\texttt{K2} = \left(\SI{12.34 (2)}{}\right)\,\SI{}{mag}$, and $M_\texttt{K1} = \left(\SI{12.56 (2)}{}\right)\,\SI{}{mag}$.
The resulting CMD is shown in Fig.~\ref{figure_cmd} and indicates that HIP~99770~b is compatible with a late L- to early T-type object.
As a consistency check, we also folded the GRAVITY spectrum with the \texttt{MKO/NSFCam.Ks} filter profile and obtained an absolute magnitude of \SI{12.544 (13)}{}, which lies within the \SI{68}{\percent} confidence interval of the discovery paper measurement of \SI{12.61 (9)}{} \citep{currie_hip99770b}.

\begin{figure}
        \centering
        \includegraphics[width=0.97\columnwidth]{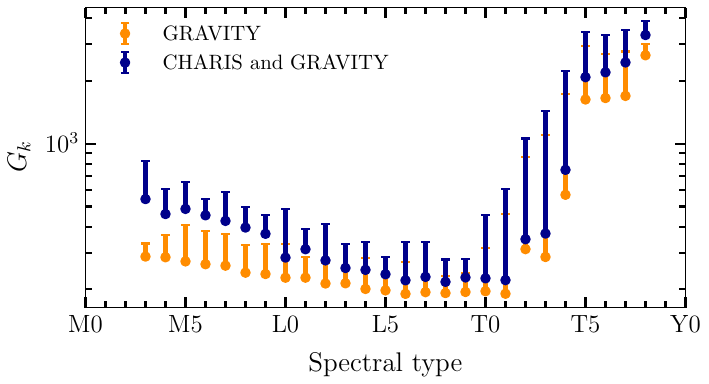}
        \caption{Goodness-of-fit statistic, $G_k$, \citep{cushing_atmospheric} as a function of spectral type when comparing the CHARIS and GRAVITY spectra together and the GRAVITY spectrum alone to empirical spectra. The minimum $G_k$ value in each spectral type bin is indicated by the coloured circles, the error bars display the full $G_k$ range resulting from all empirical spectra within the respective bin.
        }
        \label{figure_spec_type_fitting}
\end{figure}

To achieve a more quantitative classification, we used the empirical spectral type fitting routine implemented in \texttt{species}. Based on minimising the goodness-of-fit statistic, $G_k$ \citep{cushing_atmospheric}, this procedure compares the spectrum of a given object to near-infrared reference spectra of low-mass stars, brown dwarfs, and exoplanets within the SpeX Prism Library. We first performed this fit on the composite spectrum, consisting of the GRAVITY and CHARIS spectra. The latter was obtained with the original discovery of the companion \citep{currie_hip99770b}.
To ensure that the two spectra were consistent with one another, we computed a GRAVITY scaling factor using the CHARIS spectrum as a calibration baseline. Rebinning the GRAVITY spectrum onto the CHARIS wavelength solution using \texttt{SpectRes} \citep{carnall_spectres} allowed us to directly compare the two spectra in the wavelength region in which they overlap (\SI{2.00}{} to \SI{2.37}{\micro m}). We applied a simple scaling factor, $\alpha$, to the GRAVITY spectrum and subsequently compared it to the CHARIS spectrum by minimising a mutual chi-squared metric. This suggested a scaling factor of $\alpha = \SI{0.86}{}$ to facilitate the best agreement.
Fig.~\ref{figure_spec_type_fitting} shows that for the composite CHARIS and GRAVITY spectrum, the $G_k$-minimum is reached at a spectral type of L8.
To explore the effect of only having access to a narrower wavelength coverage, we performed another spectral classification run that only considered the higher-resolution GRAVITY spectrum.
In addition to preferring the earlier spectral type of L6, the GRAVITY K-band spectrum by itself proved less capable of excluding other, especially earlier spectral types. Whereas in the GRAVITY-only run, these types showed an only minimally higher and thus only marginally less appropriate goodness-of-fit statistic than the L8 bin, they were significantly disfavoured in the composite CHARIS and GRAVITY run.
As a consistency check, we repeated the spectral classification using the SpeX Prism Library Analysis Toolkit (SPLAT; \citealt{burgasser_splat}), which likewise yielded L8 for CHARIS and GRAVITY data and L6 for the GRAVITY spectrum alone.

\subsection{Constraining the system age}
\label{subsection_constraining_age}
The companion was shown to correspond to a late L- to early T-type in Sect.~\ref{subsection_spectral_classification}, which implies a cloudy atmosphere. This circumstance needs to be accounted for when selecting an appropriate evolutionary model. As is evident from Fig.~\ref{figure_cmd}, based on their physics and cloud prescriptions, the models can vary in their applicability to different objects. While the AMES-Cond \citep{allard2001limiting, baraffe2003evolutionary} and \texttt{ATMO} \citep{phillips_atmo} models appear to be poorly suited to describing objects just above the L-T transition, the AMES-Dusty model \citep{chabrier2000evolutionary, allard2001limiting} captures the position of HIP~99770~b on the CMD reasonably well.
It is thus a suitable model for constraining the system age. To this end, we used the companion Ks-band filter magnitude of $\left(\SI{12.460 (14)}{}\right)\,\SI{}{mag}$ obtained by folding the GRAVITY K-band spectrum with the \texttt{IRDIS\_B\_Ks} filter profile.
This approach clearly yielded an unrealistically precise magnitude since merely propagating the errors on the flux measurements in each wavelength channel did not take the calibration offset into account that was determined relative to the CHARIS spectrum in Sect.~\ref{subsection_spectral_classification}. Based on the scaling factor $\alpha = \SI{0.86}{}$, we adopted a systematic error of \SI{14}{\percent} which (at an \texttt{IRDIS\_B\_Ks} filter magnitude of \SI{12.46}{mag}) translates into an error of \SI{0.15}{mag}.

\begin{figure}
        \centering
        \includegraphics[width=0.95\columnwidth]{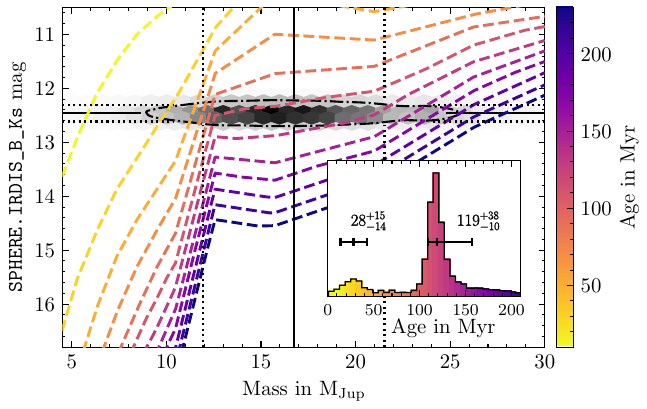}
        \caption{Isochrones from the AMES-Dusty model grid \citep{chabrier2000evolutionary, allard2001limiting} in the mass-magnitude plane. The colours indicate the respective age. The solid black lines show the dynamical mass  of the companion and the \texttt{IRDIS\_B\_Ks} filter magnitude, and the dotted lines delineate the 16th and 84th percentiles. For the filter magnitude, we adopted a conservative systematic error (see text).
        The grey colour map in the background visualises the two-dimensional distribution of randomly drawn masses and magnitudes from their respective distributions. The dash-dotted line encircles \SI{68}{\percent} of all draws. 
        The shoulder in the isochrones in this particular region of the mass-magnitude plane is caused by the onset of deuterium burning and the mass dependence of its time evolution.
        The panel in the bottom right corner shows the companion age posterior obtained by bootstrapping the drawn mass and magnitude pairs onto the age manifold as defined by the isochrones.
        }
        \label{figure_age_plot}
\end{figure}

Figure~\ref{figure_age_plot} places this filter magnitude in context with a set of AMES-Dusty grid isochrones in the vicinity of the dynamical mass of the companion. For any given mass-magnitude pair, we can compute the corresponding age as implied by the model by minimising a loss function that compares the measured magnitude to a given AMES-Dusty isochrone magnitude at the mass in question. By thus finding the isochrone that best describes the mass-magnitude pair, we obtained the best-fitting age.
We applied this bootstrapping procedure to \SI{e6}{} mass-magnitude pairs drawn from the dynamical mass posterior distribution and a Gaussian \texttt{IRDIS\_B\_Ks} distribution, $\mathcal{N}(\SI{12.46}{mag}, \SI{0.15}{mag})$, which takes the dominant systematic calibration error into account. This yielded the age distribution visualised in the small panel in the bottom right corner of Fig.~\ref{figure_age_plot}.

The bimodal nature of the age distribution is a direct consequence of the circumstance that HIP~99770~b occupies a region in the mass-magnitude plane in which deuterium can affect the cooling of the companion as it ages. 
The mass-dependent position of the deuterium-burning shoulder in the cooling (luminosity) curves at constant mass as a function of time (e.g.\ \citealt{burrows_bd_theory,saumon_evolution_l_and_t_dwarfs}; see a similar feature in \citealp{hinkley_hd206893c}) causes a ripple, or a peak\footnote{Not to be confused with a deuterium flash, which is a sudden rise, on a logarithmic scale, in the luminosity of an object that begins to burn deuterium only late (e.g.\,\citealp{bodenheimer_d_burning,marleau_sinit}).}, in the isochrones that is clearly visible in Fig.~\ref{figure_age_plot}.

Computation of the median and \SI{68}{\percent} confidence intervals of both modes yielded the two possible age scenarios of $a_\mathrm{y} = 28_{-14}^{+15}\,\mathrm{Myr}$ and $a_\mathrm{o} = 119_{-10}^{+37}\,\mathrm{Myr}$, where y and o denote younger and older, respectively.
The older scenario is strongly preferred in terms of its statistical power. There is also a non-zero probability that the companion occupies a position between the two modes in close proximity to the deuterium shoulder (see also Fig.~8 of \citealt{burrows_bd_theory}).
Using \texttt{ATMO} \citep{phillips_atmo} as the underlying evolutionary model to convert the companion K-band magnitude and dynamical mass into an age, we found similar results. The two modes are located at $a^\texttt{ATMO}_\mathrm{y} = 23_{-11}^{+20}\,\mathrm{Myr}$ and $a^\texttt{ATMO}_\mathrm{o} = 96_{-13}^{+36}\,\mathrm{Myr}$.
\citet{currie_hip99770b} placed the age of this system between \SI{40}{} and \SI{400}{Myr}, but emphasised that a self-consistent treatment similar to what we outlined above suggests an age between \SI{115}{} and \SI{200}{Myr}, which broadly agrees with our older scenario.
More precise knowledge of the companion mass would disambiguate between the two modes of our age distribution.
Thus, the main hindrance preventing a firmer age constraint is again the looseness of the dynamical mass constraint.

The above analysis and age constraint hinges on the validity of the evolutionary model for HIP~99770~b. Several aspects warrant caution. Firstly, instances in which the evolutionary models deviate from the observed characteristics of exoplanets are well documented, an example being the tensions exhibited by older substellar companions with constrained dynamical masses (e.g.\,\citealt{dupuy2009dynamical, kuzuhara2022directimaging, franson2023hip21152B}). Secondly, and perhaps more importantly for HIP~99770~b, for young objects, an uncertainty persists as to whether the onset of companion formation is simultaneous or delayed relative to the disc and host formation (e.g.\,\citealt{franson2023aflepb, zhang2023elemental, balmer2025aflepb}). Because especially at young ages, evolutionary models imply strong gradients in the bulk parameters of companions as a function of time, erroneous assumptions regarding the onset of companion formation can result in flawed inferences.

\subsection{Atmospheric forward modelling}
\label{subsection_forward_modelling}
Using \texttt{species}, we performed a suite of atmospheric grid-model fits to the entirety of the spectral and photometric data available for HIP~99770~b. As was the case for the orbit sampling procedure in Sect.~\ref{section_orbital_analysis}, every sampling run described below was performed with an underlying Gaussian parallax prior, $\mathcal{N}(\SI{24.55}{mas}, \SI{0.09}{mas})$, which corresponds to the host parallax angle as listed in \textit{Gaia} DR3 \citep{gaia_dr3}.
In all fitting procedures, each wavelength channel across both spectra was weighted equally.
Additionally, we accounted for correlations between the GRAVITY wavelength channels using the covariance matrix provided by the data reduction pipeline.
According to \citet{currie_hip99770b}, the covariance matrix associated with the CHARIS spectrum is dominated by
spatially and spectrally uncorrelated noise, which likely renders its inclusion insignificant.
Similar to the approach outlined in Sect.~\ref{subsection_spectral_classification}, we included an additional factor, $\alpha$, as a free parameter to allow for a relative flux calibration between the CHARIS and GRAVITY spectra.

\begin{figure*}
        \centering
        \includegraphics[width=0.9\textwidth]{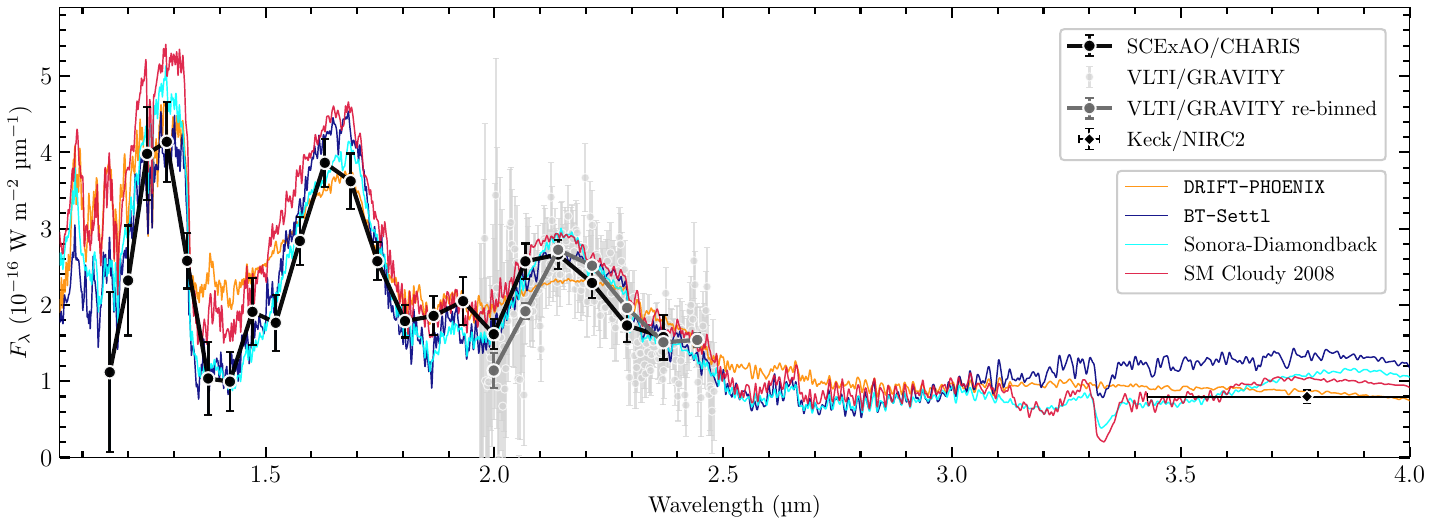}
        \caption{
        Spectral energy distribution of HIP~99770~b. The black points encircled by a white border and connected by a black line show the CHARIS spectrum taken from \citet{currie_hip99770b}. The light grey points depict the combination of the on- and off-axis GRAVITY spectra. For visualisation purposes, a rebinned lower resolution GRAVITY spectrum is shown in dark grey encircled by a white border and connected by a dark grey line. Additionally, the mid-infrared photometry data point in the \texttt{NIRC2.Lp} filter is shown as a black diamond encircled by a white border \citep{currie_hip99770b}. The associated horizontal error bar indicates the FWHM of the filter transmission profile.
        The spectra resulting from the most likely parameter sets determined by the posterior sampling processes are shown for four different atmospheric forward models. SM Cloudy 2008 denotes the cloudy ($\fsed=2$) \citet{saumon_evolution_l_and_t_dwarfs} model.
        }
        \label{figure_spectral_fits}
\end{figure*}

Cloudless models such as Sonora Bobcat \citep{marley_sonora} or Saumon \& Marley Clear (2008) \citep{saumon_evolution_l_and_t_dwarfs} did not result in adequate fits. Instead, these models tended to overestimate the spectral energy distribution below \SI{1.9}{\micro as}. This was especially the case for the two prominent broad-band spectral features surrounding the water absorption signature at \SI{1.4}{\micro as}.
The muted nature of these features suggests a cloudy atmosphere. We therefore proceeded to fit the spectral energy distribution using a set of cloudy models.
The full corner plots showing the posteriors resulting from sampling linearly interpolated grids of the \texttt{DRIFT-PHOENIX} \citep{woitke_helling_dust_ii, helling_woitke_dust_v, helling_consistent}, \texttt{BT-Settl} \citep{allard_homeier_freytag_models},
Sonora Diamonback \citep{morley_diamondback},
and Saumon \& Marley Cloudy (2008) \citep{saumon_evolution_l_and_t_dwarfs} models using \SI{1000}{} live points can be found in Appendix~\ref{app_section_additional_spectral_fitting_plots}.
As compared to the dynamical mass obtained from the refined orbital solution in Sect.~\ref{section_orbital_analysis}, the \texttt{BT-Settl} and Sonora Diamonback models performed poorly in that they inferred significantly lower masses. This is a cause for concern.
We therefore conducted an additional posterior sampling run, this time employing a Gaussian companion mass prior, $\mathcal{N}(\SI{17}{M_{Jup}}, \SI{6}{M_{Jup}})$, that reflects the dynamical mass constraint obtained in Sect.~\ref{section_orbital_analysis}. The posteriors resulting from this informative mass prior are shown alongside the initial non-informative, that is, uniform, mass prior run in Appendix~\ref{app_section_additional_spectral_fitting_plots}. For the problematic models, the inclusion of the informative mass prior was evidently not effective in forcing the walkers to sample a solution within the dynamical mass constraint.
Both the Sonora Diamondback and the \texttt{BT-Settl} posterior samplings showed little to no reaction to the application of the informative prior.
We conclude that the dynamical mass constraint is too loose to effect a strong and physically accurate pull on the posteriors. Subsequent tests with artificially reduced dynamical mass uncertainties, and thus, stronger, more informative mass priors indeed resulted in the desired forcing confirming that the low current precision of the prior is the cause for its futility.
The \texttt{DRIFT-PHOENIX} and Saumon \& Marley Cloudy (2008) models, on the other hand, yielded reasonable posterior masses in both the non-informative and informative companion mass prior cases.
To assess the performance of the different models and the inferred results on the basis of a consistent set of priors, we only consider the informative mass prior sampling runs hereafter.
Figure~\ref{figure_spectral_fits} shows the spectra associated with the most likely parameter sets for each of the chosen models. The median posterior parameter values and \SI{68}{\percent} confidence intervals are presented in Table~\ref{table_spectral_posteriors}.
The reduced chi-squared values, $\chi_\mathrm{red}^2$, of these best-fitting parameter sets also to be found in Table~\ref{table_spectral_posteriors} are acceptable for all models. This quantitatively confirms the qualitative impression that the models manage to capture the overall shape of the spectral energy distribution in Fig.~\ref{figure_spectral_fits}.
However, the model posteriors in Table~\ref{table_spectral_posteriors} and Appendix~\ref{app_section_additional_spectral_fitting_plots} reveal that the inferred parameter values can differ significantly between the individual models.

The median effective temperatures of the companion are spread across a range spanning from approximately \SI{1300}{} to \SI{1770}{K} with the Sonora Diamondback and \texttt{DRIFT-PHOENIX} models yielding the lowest and highest temperatures, respectively.
The surface gravity also varies significantly, with the Sonora Diamondback chain converging on the edge of the prior range at $\log(g) = \SI{3.6}{}$, while \texttt{DRIFT-PHOENIX} places it at approximately \SI{5.0}{}.
While \texttt{BT-Settl} and Sonora Diamondback agree on a radius of approximately \SI{1}{R_{Jup}}, Saumon Marley and \texttt{DRIFT-PHOENIX} converge on \SI{1.5}{} and \SI{0.6}{R_{Jup}}, respectively.
The flux scaling factor, $\alpha$, which is required to reconcile the GRAVITY with the CHARIS spectrum and was derived in Sect.~\ref{subsection_constraining_age}, is recovered to within $\SI{2}{\sigma}$ by all fits.
Finally, sampling the companion atmosphere metallicity is only supported by the \texttt{DRIFT-PHOENIX} and Sonora Diamondback models.
With an inferred metallicity of $[\mathrm{Fe}/\mathrm{H}]=0.25_{-0.06}^{+0.03}$, the chain converges close to the edge of the \texttt{DRIFT-PHOENIX} grid at \SI{0.3}{}. The value agrees well with the somewhat looser $[\mathrm{Fe}/\mathrm{H}]=0.27_{-0.12}^{+0.12}$ obtained using the Sonora Diamondback model, however, which is defined over a wider grid that extends to \SI{0.5}{}. Furthermore, they are both consistent with the metallicity measurement of $[\mathrm{M}/\mathrm{H}]=0.26_{-0.23}^{+0.24}$ reported by \citet{zhang_hip99770b}.
An extension of the spectral coverage into the mid-infrared, for instance by means of JWST observations, is likely to further constrain the posteriors, and would thus enable a tighter hold on the companion metallicity and elemental abundance ratios (e.g.\ see \citealt{miles2023ers}).

\begin{figure}
        \centering
        \includegraphics[width=0.98\columnwidth]{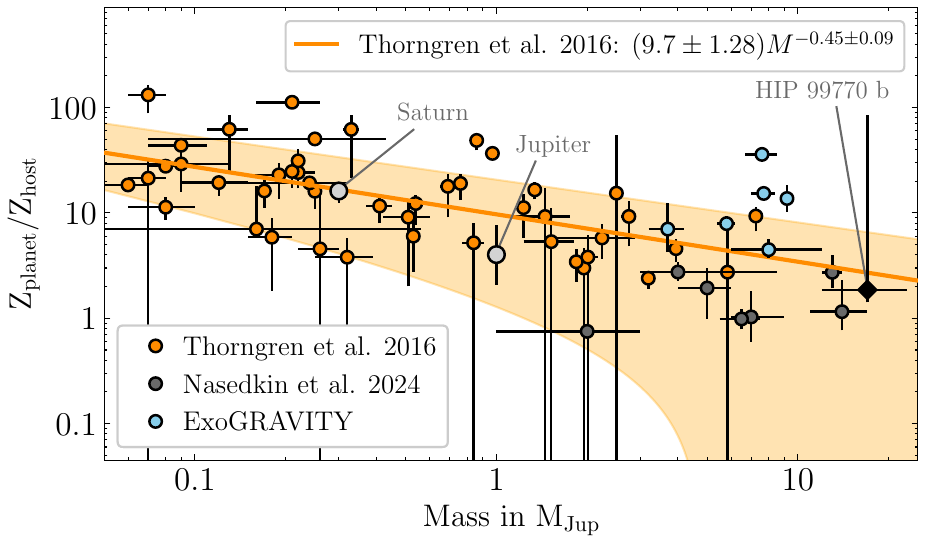}
        \caption{
        Mass-metallicity plane populated with companions from different samples, adapted from \citet{thorngren2016mass}, \citet{nasedkin2024four} and \citet{balmer2024hd136164Ab}. The orange points show transiting giant exoplanets from \citet{thorngren2016mass}. The grey points illustrate a set of directly detected planets, and the blue points show a subset that was observed using VLTI/GRAVITY. Both samples were compiled by \citet{nasedkin2024four} and references therein. For context, Jupiter and Saturn are included with the metallicity measurements taken from \citet{guillot1999comparison}. The orange line and confidence interval trace the empirical relation between mass and metallicity for giant exoplanets presented by \citet{thorngren2016mass}. The black diamond indicates the ratio derived for HIP~99770~b when assuming a solar metallicity for the host. We took the higher metallicity ratios obtained from the host metallicity estimates by \citet{paunzen1999accurate, paunzen2002period}, and \citet{villaume2017extended} into account by presenting the data point with an inflated error bar that reflects a skewed systematic error encompassing the \SI{68}{\percent} confidence interval of the highest encountered ratio.
        }
        \label{figure_mass_metallicity_relation}
\end{figure}

Following the analysis presented by \citet{nasedkin2024four} and \citet{balmer2025aflepb}, we can contextualise this result with other planetary metallicity measurements. To this end, we converted the metallicity relative to the solar value into a metallicity ratio of the companion and the host.
There are different metallicity constraints for HIP~99770 in the literature: \citet{paunzen1999accurate} placed it at $[\mathrm{Fe}/\mathrm{H}]_\mathrm{host}=\left(\SI{-1.3 (2)}{}\right)$, \citet{paunzen2002period} estimated $\left(\SI{-1.46 (8)}{}\right)$, while \citet{villaume2017extended} inferred $\SI{-0.8}{}$ (no uncertainty reported).
These estimates, however, might be superficial since HIP~99770 has been classified as a chemically peculiar $\lambda$ Boo star \citep{murphy2017}.
Despite the apparent metal depletion suggested by their iron underabundances, these stars are in fact expected to possess bulk solar abundances (e.g.\,\citealt{murphy2020discovery}).
Since the companion metallicity inferred using \texttt{DRIFT-PHOENIX} ($[\mathrm{Fe}/\mathrm{H}]_\mathrm{comp}=0.25_{-0.06}^{+0.03}$) is based on a chain that converged towards the edge of the defined grid range, we only considered the estimate obtained using the Sonora Diamondback model ($0.27_{-0.12}^{+0.12}$).
This yielded a metallicity ratio of companion and host of $1.9_{-0.4}^{+0.6}$, which would be consistent with the mass-metallicity relation derived by \citet{thorngren2016mass}.
How these results compare to other companions with measured metallicity ratios as well as the aforementioned empirical relation is visualised in Fig.~\ref{figure_mass_metallicity_relation}.

The main caveat for assessing the derived metallicity ratios is their dependence on the loosely constrained stellar metallicity.
To illustrate, if we had instead used a host metallicity of $[\mathrm{Fe}/\mathrm{H}]_\mathrm{host}=\left(\SI{-1.46 (8)}{}\right)$, as suggested by \citet{paunzen2002period}, the metallicity ratio would have amounted to $50_{-20}^{+30}$, which would indicate strong metal enrichment in the companion.
This conclusion might be interpreted in terms of the sequence of formation phases that have led to the current state of the companion.
Indeed, the high metallicity of many giant planets as compared to their host stars suggests that they have accreted a significant amount of solid material after their initial formation process, a mechanism referred to as late accretion (e.g.\,\citealt{mousis2009determination, franson2023aflepb, zhang2023elemental, balmer2025aflepb}).

\begin{table}[t]
    \centering
    \caption{Atmospheric forward modelling posteriors.}
    \label{table_spectral_posteriors}
    \renewcommand{\arraystretch}{1.4}
    \resizebox{\columnwidth}{!}{%
    \begin{tabular}{lcccccc}
    \toprule
    Model &
    $\chi_\mathrm{red}^2$ &
    $T_\mathrm{eff}$ (K) &
    $\log(g)$ &
    $R$ ($\mathrm{R}_\mathrm{Jup}$) &
    $M$ ($\mathrm{M}_\mathrm{Jup}$) &
    $[\mathrm{Fe}/\mathrm{H}]$ \\
    \midrule
    \midrule
    \texttt{DRIFT-PHOENIX} &
    \SI{2.06}{} &
    $1772_{-11}^{+11}$ &
    $5.03_{-0.06}^{+0.08}$ &
    $0.587_{-0.009}^{+0.011}$ &
    $15_{-2}^{+3}$ &
    $0.25_{-0.06}^{+0.03}$ \\
    \texttt{BT-Settl} &
    \SI{1.25}{} &
    $1380_{-20}^{+20}$ &
    $3.95_{-0.02}^{+0.02}$ &
    $0.99_{-0.03}^{+0.04}$ &
    $3.5_{-0.2}^{+0.2}$ &
    - \\
    Sonora-Diamondback &
    \SI{1.06}{} &
    $1300_{-20}^{+20}$ &
    $3.62_{-0.08}^{+0.15}$ &
    $1.05_{-0.04}^{+0.04}$ &
    $1.8_{-0.3}^{+0.7}$ &
    $0.27_{-0.12}^{+0.12}$ \\
    SM Cloudy 2008\hyperlink{ref_smc2008}{\textsuperscript{*}} &
    \SI{2.08}{} &
    $1539_{-27}^{+17}$ &
    $4.16_{-0.06}^{+0.07}$ &
    $1.5_{-0.3}^{+0.3}$ &
    $13_{-4}^{+4}$ &
    - \\
    \bottomrule
    \end{tabular}
    }
    \tablefoot{The values were rounded to their significant figure. They correspond to the median of the marginalised posterior distribution and its differences to the 84th and 16th percentiles in superscript and subscript, respectively. These results stem from the sampling runs employing an informative underlying companion mass prior (see corner plots in Appendix~\ref{app_section_additional_spectral_fitting_plots} for uninformative mass prior posteriors). \\
    \hypertarget{ref_smc2008}{\textsuperscript{*}} $\fsed=2$ model of \citet{saumon_evolution_l_and_t_dwarfs}.
    }
\end{table}

\subsection{Comparing evolutionary and atmospheric modelling}
\label{subsection_atmo_evo_comparison}

\begin{figure}
        \centering
        \includegraphics[width=0.9\columnwidth]{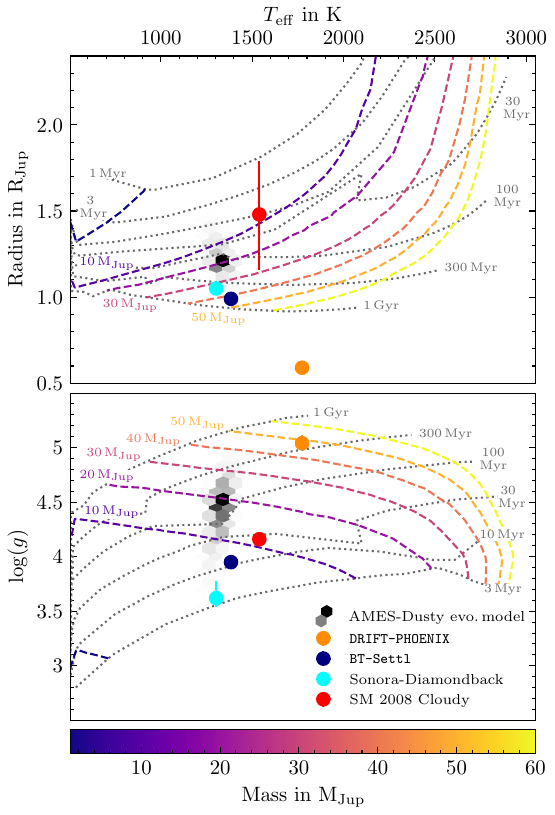}
        \caption{
        Companion radius, $R$, and surface gravity, $\log(g)$, as a function of effective temperature, $T_\mathrm{eff}$. The dotted grey lines indicate isochrones of different ages, the dashed lines illustrate evolutionary tracks, i.e. how a companion of a certain mass evolves over time. The specific masses that the tracks correspond to are indicated by their respective colours. Both the isochrones and the evolutionary tracks are taken from the AMES-Dusty model grid \citep{chabrier2000evolutionary, allard2001limiting}.
        The hexagonal bin map in the background shows where a sample of pairs drawn from the mass and K-band magnitude distributions falls when propagated into the parameter planes depicted in the two respective panels using AMES-Dusty.
        Finally, the values obtained through the atmospheric models applied to the spectral energy distribution of the companion are marked as circles of different colours.
        }
        \label{figure_evo_atmo_comparison}
\end{figure}

To assess whether the bulk parameter values inferred using atmospheric models to fit the spectral energy distribution as discussed in Sect.~\ref{subsection_forward_modelling} are physically plausible, we compared them to the values implied by self-consistent evolutionary models. To this end, the relations between the companion effective temperature, $T_\mathrm{eff}$, and its radius, $R$, as well as between its effective temperature and surface gravity, $\log(g)$, at constant ages (isochrones) and constant masses (evolutionary tracks) as suggested by the AMES-Dusty model are plotted alongside the respective values resulting from sampling the different atmospheric model grids in Fig.~\ref{figure_evo_atmo_comparison}. We also highlight the $T_\mathrm{eff}$-$R$ and $T_\mathrm{eff}$-$\log(g)$ regions consistent with the AMES-Dusty model at the dynamical mass and age of HIP~99770~b (via its Ks-band magnitude, as explained in Sect.~\ref{subsection_constraining_age}).

Comparison of these self-consistently inferred parameter regions with the values from the atmospheric modelling implies a poor agreement between the two independent methods. Noticeably, the \texttt{DRIFT-PHOENIX} results are farthest removed from those of AMES-Dusty in both cases. Fig.~\ref{figure_evo_atmo_comparison} reveals that the \texttt{DRIFT-PHOENIX} radius of approximately \SI{0.6}{R_{Jup}} is unrealistic at the ages suggested by both the older and the younger scenario outlined in Sect.~\ref{subsection_constraining_age}. Similarly, the surface gravity inferred by \texttt{DRIFT-PHOENIX} would necessitate a far more massive body, which we know to be impossible from the orbital analysis in Sect.~\ref{section_orbital_analysis}.
The effective temperature is reasonably well recovered by the three remaining atmospheric models. However, the radius and surface gravity values from the atmospheric models both show a significant degree of scatter around the feasible parameter regions implied by AMES-Dusty.
Seeing that it overlaps with the AMES-Dusty model in both panels, of the three, the Saumon \& Marley Cloudy (2008) model performs best, while \texttt{BT-Settl} and Sonora-Diamondback underestimate both radius and surface gravity.
Thus, even in cases where the dynamical mass is only loosely constrained around the deuterium-burning threshold, the radius, surface gravity, and effective temperature of the companion as derived from self-consistent evolutionary models still exhibit less scatter than the scatter that is encountered between different atmospheric models.

Going forward, the CHARIS, GRAVITY and NIRC2 observations at hand constitute a suitable data set for testing and gauging atmospheric models in comparison to self-consistent evolutionary models.

\section{Conclusions}
\label{section_conclusion}

Here, we have presented an updated study of the directly imaged super-Jupiter HIP~99770~b on the basis of two new data sets obtained by the near-infrared interferometric instrument VLTI/GRAVITY.
These additional observations added a highly precise astrometric epoch of the companion position relative to its host and extended extended the temporal coverage available for the system. In addition to confirming the results reported by \citet{currie_hip99770b}, the combination of these gains also served to further constrain the orbital solution of the companion.
Due to the large relative uncertainty of the host PMa, the resulting companion dynamical mass constraint is still comparatively loose at $17_{-5}^{+6}\,\mathrm{M}_\mathrm{Jup}$.
We showed that this situation cannot be remedied by the addition of further relative astrometric epochs.
Thus, when its relative uncertainty is large, the dominant nature of the PMa in the orbital sampling procedure prevents further constraints on the dynamical mass of the companion.
The only viable method of gaining a firmer grasp on it in the foreseeable future is to exploit the time-series astrometry to be published in \textit{Gaia} DR4.

While it was unable to strongly constrain the companion mass, the orbital resampling resulted in a significant detection of a non-vanishing eccentricity at $0.31_{-0.12}^{+0.06}$.
Although this moderately elevated eccentricity was conceivable even before inclusion of the GRAVITY epoch, the CHARIS and NIRC2 epochs by themselves were also consistent with a low or even vanishing orbital eccentricity, which is now positively ruled out.
Instead, HIP~99770~b appears more eccentric than most directly imaged exoplanets.

Comparing the GRAVITY K-band magnitude and dynamical mass of the companion to an evolutionary model, we found them to be consistent with two scenarios. The first implies a planetary mass object at an age of $28_{-14}^{+15}\,\mathrm{Myr}$, the second suggests a more massive object beyond the deuterium-burning threshold at an age of $119_{-10}^{+37}\,\mathrm{Myr}$.
The full spectral energy distribution of the companion is best described by an empirical L8 type object. This agrees with the location of the body on a CMD as compared to a literature population of brown dwarfs and exoplanets.

Next, we fitted the observed spectrum of the companion using different atmospheric models. While each model performed well in terms of its reduced $\chi^2$ squared value, the inferred parameter values varied significantly. The use of an informative mass prior based on the dynamical mass of the companion was implemented, but proved ineffective due to its comparative looseness. Employing a firmer dynamical mass constraint as a prior would effect a stronger pull on the posteriors.
The inferred enriched metallicity of the companion relative to the solar value is consistent across different models and with the results presented by \citet{zhang_hip99770b}.

Finally, the radii, surface gravities, and effective temperatures inferred from the atmospheric models were compared to the results obtained from a self-consistent evolutionary model.
This approach revealed the \texttt{DRIFT-PHOENIX} results to be inconsistent with companion's mass and age for both the younger and older scenario outlined above.
While the other models yielded values more in line with the evolutionary approach, this agreement is feeble and we encountered considerable scatter between the models.
Despite its current looseness, the dynamical mass of the companion reveals significant incongruities between self-consistent evolutionary and atmospheric models. Not only do the latter infer numerical parameter values that are at odds with the known physics of substellar companions, they are also inconsistent with each other.

The results laid out in this work do not permit us to infer the formation history of HIP~99770~b. That being said, the provided astrometric and spectroscopic data as well as the new constraints they facilitated will support endeavours to do so in the future.


\begin{acknowledgements}
This work has made use of data from the European Space Agency (ESA) mission
{\it Gaia} (\url{https://www.cosmos.esa.int/gaia}), processed by the {\it Gaia}
Data Processing and Analysis Consortium (DPAC,
\url{https://www.cosmos.esa.int/web/gaia/dpac/consortium}). Funding for the DPAC
has been provided by national institutions, in particular the institutions
participating in the {\it Gaia} Multilateral Agreement.
This research has made use of the Jean-Marie Mariotti Center \texttt{Aspro}
service \footnote{Available at http://www.jmmc.fr/aspro}.
This research has benefitted from the SpeX Prism Spectral Libraries, maintained by Adam Burgasser at\url{http://www.browndwarfs.org/spexprism}.
S.L.\ acknowledges the support of the French Agence Nationale de la Recherche (ANR-21-CE31-0017, ExoVLTI) and of the European Union (ERC Advanced Grant 101142746, PLANETES).
G.-D.M.\ acknowledges the support from the European Research Council (ERC) under the Horizon 2020 Framework Program via the ERC Advanced Grant ``ORIGINS'' (PI: Henning), Nr.~832428,
and via the research and innovation programme ``PROTOPLANETS'', grant agreement Nr.~101002188 (PI: Benisty). \\
Finally, we would like to thank the anonymous referee whose thorough feedback helped improve the paper.

\end{acknowledgements}


%
%




\bibliographystyle{aa} 
\bibliography{refs.bib} 

\begin{appendix}
\onecolumn

\section{Additional orbital fitting plots}
\label{app_section_additional_orbital_fitting_plots}
\begin{figure*}[h!]
        \centering
        \includegraphics[width=0.99\textwidth]{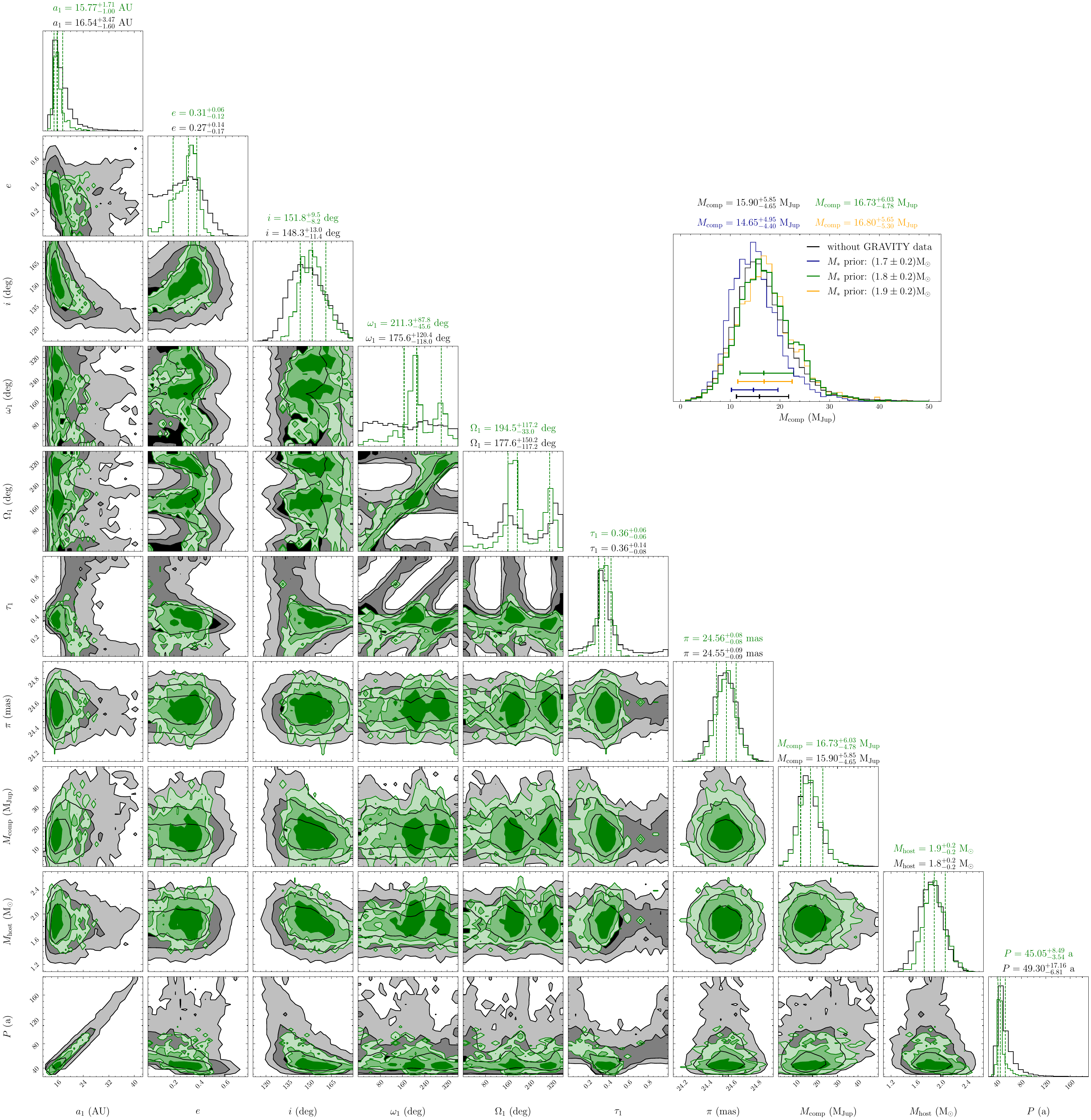}
        \caption{Corner plot showing the orbital element posteriors of HIP~99770~b. The sampling resulting from only accounting for the previously available data and that obtained upon inclusion of the GRAVITY astrometric epoch (on-axis observation on 31 May 2023) are shown in black and green, respectively. The reported values above the panels showing the marginalised posterior distributions correspond to the median and boundaries of the \SI{68}{\percent} confidence intervals of the respective element in each sampling run. The top right panel shows companion mass posterior distributions sampled using Gaussian stellar mass priors whose centres were shifted from \SI{1.8}{} to \SI{1.7}{} and \SI{1.9}{M_\odot}, respectively.}
        \label{figure_full_orbit_corner_plot}
\end{figure*}

Here, we present additional plots relating to the orbital analysis outlined in Sect.~\ref{section_orbital_analysis}.
The full corner plot showing how the inclusion of the GRAVITY astrometric epoch affects the posterior sampling of the orbital solution is shown in Fig.~\ref{figure_full_orbit_corner_plot}.

\clearpage

\newpage
\section{Additional spectral fitting plots}
\label{app_section_additional_spectral_fitting_plots}
Here, we present additional plots relating to the spectral analysis outlined in Sect.~\ref{section_spectral_analysis}.

\begin{figure*}[h!]
        \centering
        \includegraphics[width=0.95 \textwidth]{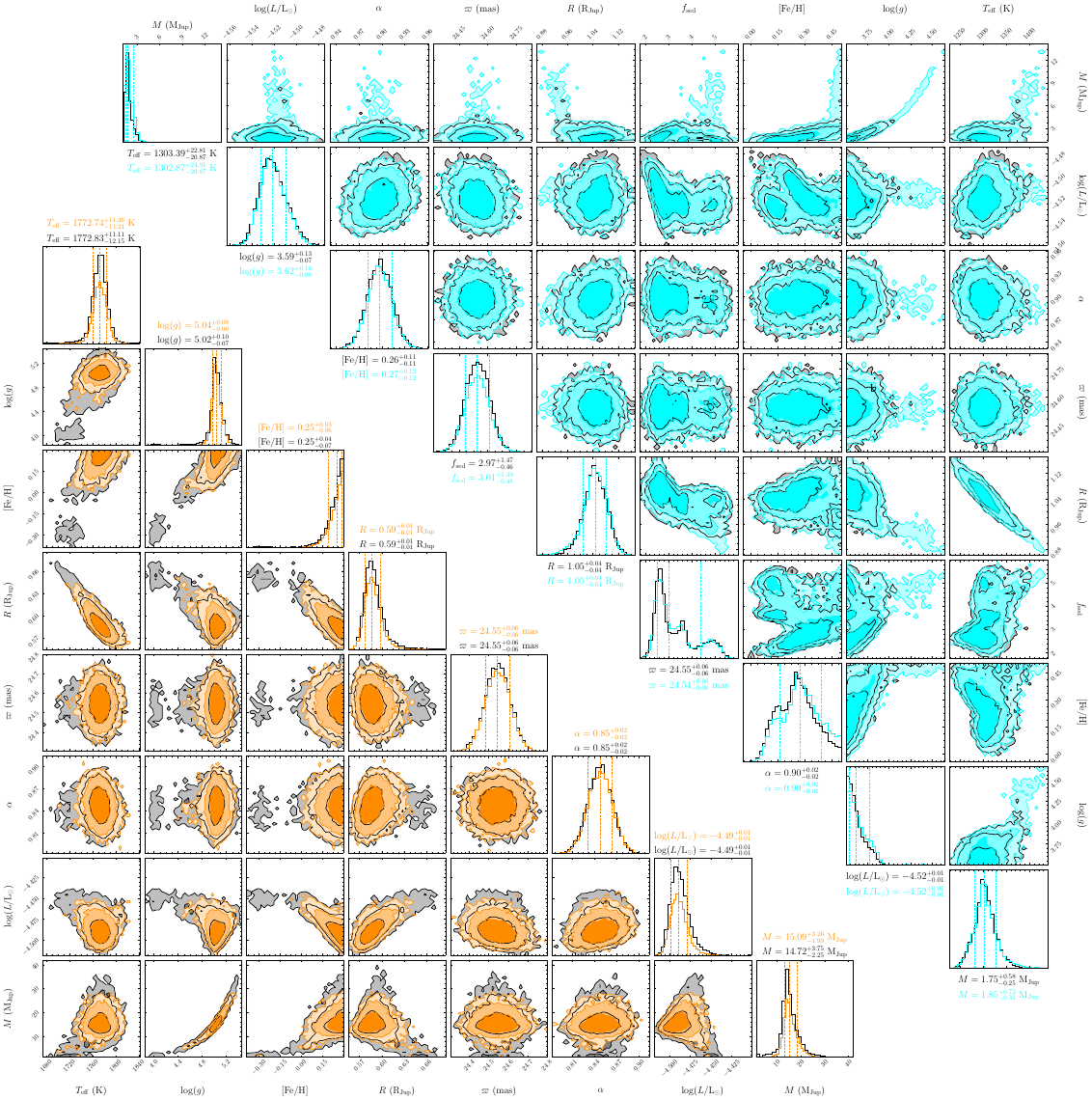}
        \caption{\textit{Lower left:} Corner plot showing the posterior sampling of the parameter grid when applying the \texttt{DRIFT-PHOENIX} model to the full set spectroscopic and photometric data presented in Sect~\ref{section_observations}. Two sampling runs were performed: the results obtained when using no mass prior, that is an uninformative uniform prior, and when using a Gaussian prior based on the dynamical mass obtained from the orbital fit (see Sect.~\ref{section_orbital_analysis}) are shown in black and orange, respectively. Above the panels showing the marginalised posterior distributions we report their median values and their differences to the 84th and 16th percentiles in superscript and subscript, respectively.
        \textit{Upper right:} Same as lower left for the Sonora Diamondback model grid.
        }
        \label{figure_ccp_drift_phoenix_sonora_diamondback}
\end{figure*}

\begin{figure*}[h!]
        \centering
        \includegraphics[width=0.95 \textwidth]{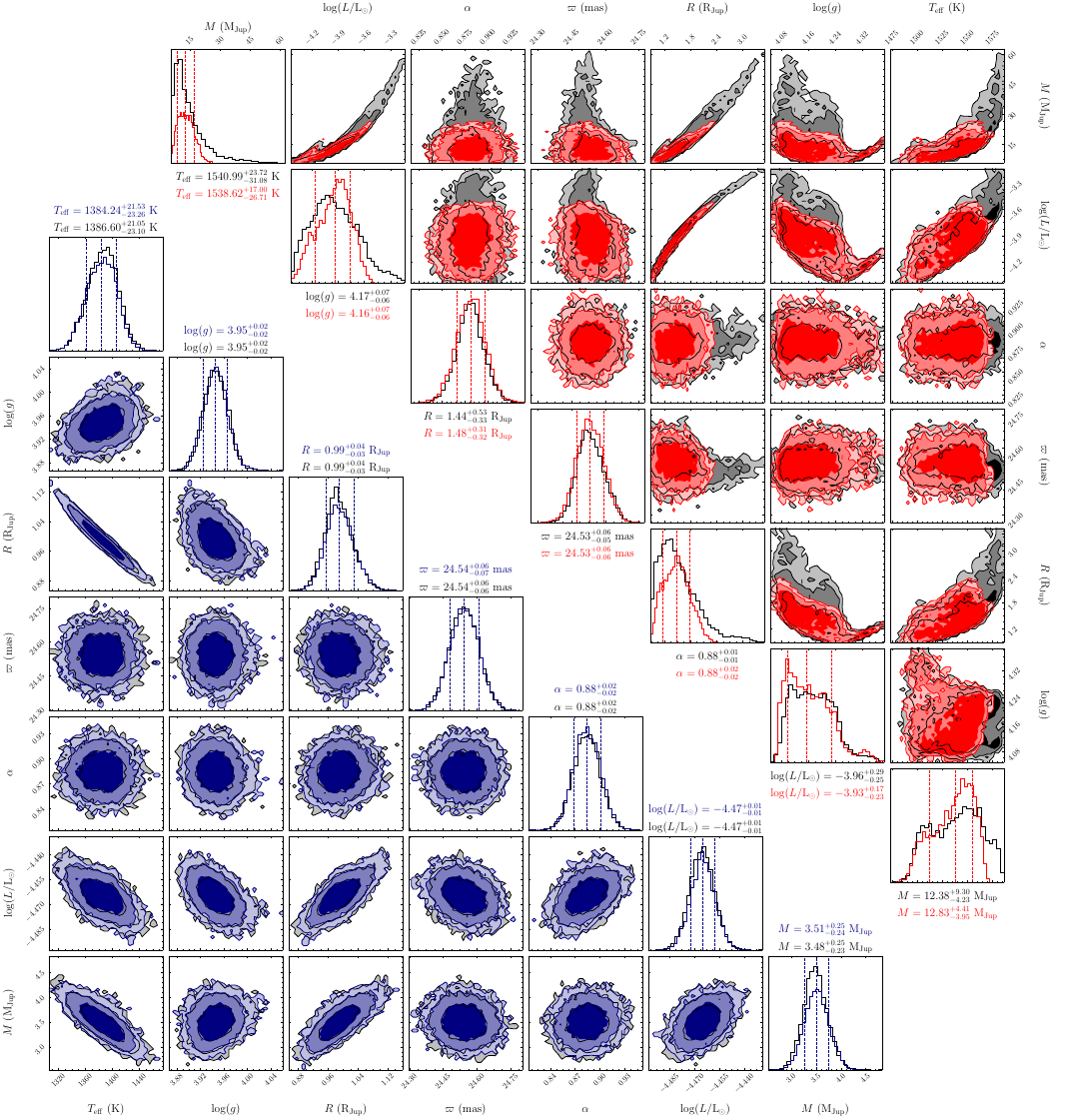}
        \caption{Same as Fig.~\ref{figure_ccp_drift_phoenix_sonora_diamondback} for the \texttt{BT-Settl} and Saumon, Marley Cloudy (2008) model grids in the lower left and upper right panels, respectively.}
        \label{figure_ccp_bt_settl_smcl2008}
\end{figure*}
\clearpage

\newpage
\section{Correcting for throughput losses}
\label{app_section_correcting_for_tp_losses}
\begin{figure}[h]
        \centering
        \includegraphics[width=0.49\columnwidth]{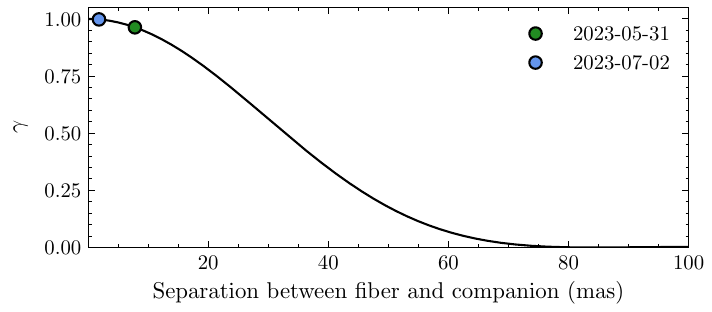}
        \caption{Normalised coupling efficiency, $\gamma$, as a function of the misplacement between the centre of the science fibre and the target. The $\gamma$-values used to correct for the throughput losses incurred during the two observations are indicated.}
        \label{figure_coupling_efficiency}
\end{figure}

Misplacement between the centre of the GRAVITY science fibre and the target aimed for (in our case the companion) leads to a loss of flux throughput when observing in dual-field mode. To correct for this, we divided the contrast spectra by the normalised coupling efficiency, $\gamma$, of the respective observations. These values can be found in Table~\ref{table_obs_log}. The coupling efficiency, visualised in Fig.~\ref{figure_coupling_efficiency}, varies between \SI{1}{} and \SI{0}{} and is a function the angular separation between the fibre centre and the target. A comprehensive derivation of its analytic description can be found in \citet{wang_constraining_nature_pds70}.

\section{Reduction of GRAVITY on and off-axis spectra}
\label{app_section_GRAVITY_spectral_reduction}
\begin{figure*}[h!]
    \centering
    \includegraphics[width=0.99\textwidth]{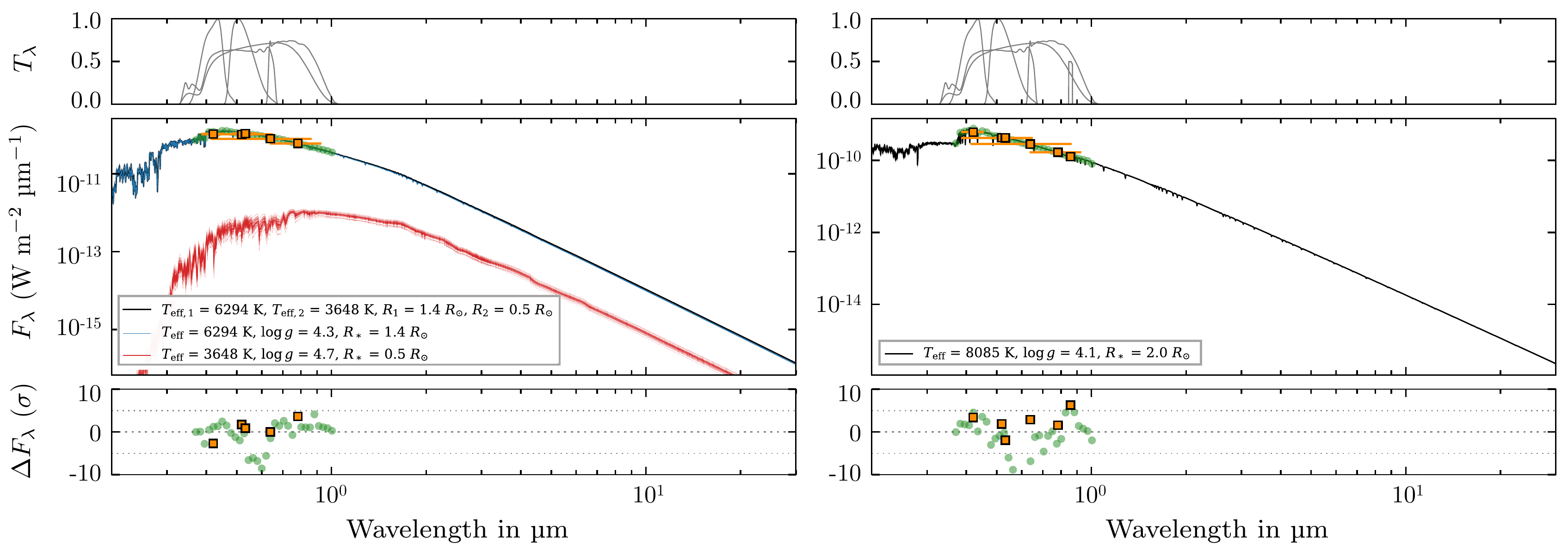}
    \caption{Inferred stellar model atmospheres for HIP~99770~A (left) and HD~196885~AB (right). The inferred model atmospheres for the A and B components are shown in blue and red, respectively. The combined model is shown in black. The thin lines illustrate 30 randomly drawn samples from the posterior distribution. The photometry and the \textit{Gaia} XP spectrum included in the fit are shown in orange and green. For greater clarity, only every 10th data point of the \textit{Gaia} XP spectrum is shown. The top panel shows the filter transmission curve for each photometric point while the bottom panel presents the residuals between the data and the best fit model atmosphere.}
    \label{figure_stellar_calibrators}
\end{figure*}

As described in Sect.~\ref{subsect_vlti_observations}, eventually arriving at the combined companion flux spectrum used in the analysis presented in Sect.~\ref{section_spectral_analysis} required converting the contrast spectra into companion flux spectra. This involved multiplying the contrast spectra by calibrator spectra that describe the reference targets (HIP~99770~A and the binary HD~196885~AB for the on and off-axis observations, respectively).
In Fig.~\ref{figure_stellar_calibrators}, we present plots visualising the fitting procedure yielding the best-fit stellar models with Table~\ref{tab:stellar_parameters} listing the parameter values associated to the highest-likelihood sample.
We did not account for a potential variability of the star. If present, such variability can impact the goodness of our fits.

These best fitting stellar models, $F_\mathrm{host}^\mathrm{model}$ for HIP~99770~A as well as $F_\mathrm{A}^\mathrm{model}$ and $F_\mathrm{B}^\mathrm{model}$ for the binary components of HD~196885~AB, are the calibrator spectra that were used to convert the observed contrast spectra, $C$, into companion flux spectra, $F_\mathrm{comp}$, for both epochs via
\begin{equation}
    \qquad \qquad \qquad \qquad \qquad
    F_\mathrm{comp,~on-axis} = C \cdot F_\mathrm{host}^\mathrm{model}, 
    \qquad \quad
    F_\mathrm{comp,~off-axis} = C \cdot \langle F_\mathrm{host}^\mathrm{FT} \rangle \cdot \sqrt{\frac{F_\mathrm{A}^\mathrm{model} \cdot F_\mathrm{B}^\mathrm{model}}{\langle F_\mathrm{A}^\mathrm{model} \rangle \cdot \langle F_\mathrm{B}^\mathrm{model} \rangle}}, \label{equation_off_axis}
\end{equation}
where the operator $\langle \rangle$ denotes the averaging over the GRAVITY bandpass. The flux average of the host star $\langle F_\mathrm{host}^\mathrm{FT} \rangle$ that is required in Eq.~\ref{equation_off_axis} is obtained from the simultaneous GRAVITY fringe tracker (FT) observations of the host at low spectral resolution. The covariances provided by the GRAVITY pipeline and stellar model uncertainties obtained by taking the standard deviation over 100 randomly drawn samples from the respective posteriors were propagated through each reduction step.

\begin{table*}[h!]
    \centering
    \caption{Inferred stellar model atmosphere parameters.}
    \label{tab:stellar_parameters}
    \begin{tabular}{ccccccccccc}
    \toprule
    Star &
    $T_\mathrm{eff, A}$ (\SI{}{K}) &
    $\log g_\mathrm{A}$ &
    $R_\mathrm{A}$ (\SI{}{R_\odot}) &
    $T_\mathrm{eff,B}$ (\SI{}{K}) &
    $\log g_\mathrm{B}$ &
    $R_\mathrm{B}$ (\SI{}{R_\odot}) &
    $\pi$ (\SI{}{mas}) &
    $A_V$ (\SI{}{mag}) &
    $f_{K_\mathrm{s}}$\\
    \midrule
    \midrule
    HIP~99770 & $8085^{+33}_{-35}$ & $4.09^{+0.05}_{-0.06}$ & $1.96^{+0.01}_{-0.01}$ & & & & $24.54^{+0.06}_{-0.07}$ & 0.043\hyperlink{ref_extinction_param}{\textsuperscript{*}} \\
    HD~196885 & $6294^{+17}_{-18}$ & $4.29^{+0.03}_{-0.03}$ & $1.36^{+0.01}_{-0.01}$ & $3648^{+129}_{-126}$ & $4.65^{+0.06}_{-0.06}$ & $0.52^{+0.04}_{-0.03}$ & $29.41^{+0.02}_{-0.02}$ & 0 & $0.059\pm0.006$\\
    \bottomrule
    \end{tabular}
    \tablefoot{
    The indices A and B denote the different binary components.
    $f_{K_\mathrm{s}}$ is the \texttt{Paranal}\texttt{/}\texttt{SPHERE}\texttt{.}\texttt{IRDIS\_B\_Ks} band contrast between the components as estimated from the GRAVITY observations. \\
    \hypertarget{ref_extinction_param}{\textsuperscript{*}} \citet{murphy2017} \\
    }
\end{table*}

\end{appendix}

\end{document}